\documentclass[journal]{IEEEtran}

\usepackage[utf8]{inputenc}

\usepackage{times,enumerate} 
\usepackage[usenames,dvipsnames,svgnames,table]{xcolor}
\usepackage{graphicx}
\usepackage[noadjust]{cite}

\usepackage{rotating}
\usepackage{csquotes}
\usepackage{amsmath}
\usepackage{bm}
\usepackage{tabularx}
\usepackage{stfloats}
\usepackage{threeparttable}
\usepackage{url}
\usepackage{soul}
\usepackage{threeparttable}
\soulregister\cite7
\soulregister\ref7
\soulregister\pageref7
\usepackage{amssymb}
\usepackage{soul}
\usepackage{footnote}
\usepackage{colortbl}
\usepackage{soul}
\usepackage{multirow}
\usepackage{pifont}
\usepackage{algorithmicx}
\usepackage{booktabs,xcolor,colortbl}
\usepackage{color}
\usepackage{alltt}
\usepackage{hyperref}
\usepackage{enumerate}
\usepackage{siunitx}
\usepackage{epstopdf}
\usepackage{bbding}
\usepackage{pbox}
 \usepackage{amsmath}  
 \usepackage{amssymb}   
 \usepackage{amsthm}
 \usepackage[dvipsnames]{xcolor}

\usepackage{algorithm}
\usepackage{algpseudocode}

\newcommand{\probP}{\text{I\kern-0.15em P}}
\newcommand{\probE}{\text{I\kern-0.15em E}}

\usepackage{nomencl}

\begin{document}

\title{
Dynamic Quantum Key Distribution for Microgrids with Distributed Error Correction}
\author{Suman Rath, Neel Kanth Kundu, \textit{Member, IEEE} and Subham Sahoo, \textit{Senior Member, IEEE} 
\thanks{

Suman Rath is with the Department of Computer Science and Engineering, University of Nevada, Reno, NV 89557, USA (e-mail: sumanr@unr.edu).}
\thanks{Neel Kanth Kundu is with the Centre for Applied Research in Electronics
(CARE), Indian Institute of Technology Delhi, New Delhi-110016, India
(e-mail: neelkanth@iitd.ac.in). He is also an honorary fellow at the
Department of Electrical and Electronic Engineering, University of
Melbourne, Melbourne, VIC, Australia. His work is supported in part by the INSPIRE
Faculty Fellowship awarded by the Department of Science and Technology,
Government of India (Reg. No.: IFA22-ENG 344), and the New Faculty Seed
Grant (NFSG) from the Indian Institute of Technology Delhi.
} 
\thanks{Subham Sahoo is with the Department of Energy, Aalborg University, 9220 Aalborg, Denmark (e-mail: sssa@energy.aau.dk).}}

\maketitle

\begin{abstract}

Quantum key distribution (QKD) has often been hailed as a reliable technology for secure communication in cyber-physical microgrids. Even though unauthorized key measurements are not possible in QKD, attempts to read them can disturb quantum states leading to mutations in the transmitted value. Further, inaccurate quantum keys can lead to erroneous decryption producing \textit{garbage} values, destabilizing microgrid operation. QKD can also be vulnerable to node-level manipulations incorporating attack values into measurements before they are encrypted at the communication layer.
To address these issues, this paper proposes a secure QKD protocol that can identify errors in keys and/or nodal measurements by observing violations in control dynamics. Additionally, the protocol uses a dynamic adjacency matrix-based formulation strategy enabling the affected nodes to reconstruct a trustworthy signal and replace it with the attacked signal in a multi-hop manner. This enables microgrids to perform nominal operations in the presence of adversaries who try to eavesdrop on the system causing an increase in the quantum bit error rate (QBER).
We provide several case studies to showcase the robustness of the proposed strategy against eavesdroppers and node manipulations. The results demonstrate that it can resist unwanted observation and attack vectors that manipulate signals before encryption.

\end{abstract}


\IEEEpeerreviewmaketitle

\section{Introduction}

\IEEEPARstart{M}{icrogrids} are small-scale energy systems that facilitate local energy production and integration of renewable energy sources, to help realize a carbon-neutral power sector. Such systems can operate in both, autonomous and grid-connected modes. Parameter regulations in microgrids can be achieved via a centralized or distributed hierarchical control structure.
Supervisory control and data acquisition (SCADA)-enabled data aggregation and processing in centralized microgrids can expose the system to single-point failure \cite{sahoo2019cyber}. On the other hand, distributed systems can be more challenging to manipulate as the data processing burden is spread over several localized servers, leading to higher reliability. 
Hence, distributed control is more widely prevalent in cyber-physical microgrids. It also enables increased tolerance against issues like communication failures, packet drops, and propagation delays \cite{rath2020cyber}. 
Despite several advantages, distributed microgrid control can also have some limitations. This is mainly due to the possible presence of cyber-attackers with illegitimate access to sensors, actuators, and/or network devices within the microgrid environment \cite{sahoo2018stealth}. As microgrids are being deployed in mission-critical domains like military bases \cite{booth2010targeting}, hospitals, and electric aircraft/ships, cyber-attacks can have a devastating impact on national security and society.

Distributed microgrids can be vulnerable to eavesdropping and packet injection attacks as they rely on communication modules to achieve coordination between the scattered local servers in the network. Hence, ensuring the integrity of data flowing through communication links is essential. One of the ways to achieve reliability for such data packets is through the use of encryption and time stamps \cite{rath2020cyber}. An Advanced Encryption Scheme (AES)-based cryptographic scheme was proposed in \cite{iqbal2018low} to enhance privacy in SCADA-based microgrid networks. Even though AES is a widely accepted standard of encryption in cyber-physical systems, it may be vulnerable to side-channel attacks \cite{Basatwar_2023}. Side-channel attacks refer to the exploitation of information unintentionally leaked by a physical cryptosystem or device during nominal operations \cite{lawson2009side}. Instead of directly targeting the cryptographic algorithms or security mechanisms themselves, side-channel attacks focus on analyzing the physical characteristics or implementation details of the system to gather sensitive information. Apart from side-channel attacks, AES can also be vulnerable to man-in-the-middle and brute-force attacks. It is important to protect microgrid networks against brute-force attacks as they form a significant proportion of typical network attacks and can have a catastrophic effect on system stability \cite{salamatian2019botnets}. Chen \textit{et al.} \cite{chen2022privacy} proposed a homomorphically encrypted consensus algorithm that allows secure communication among microgrid entities without a third party. In addition to this, the authors also proposed an estimator-like dynamic quantizer to facilitate data encryption, ensuring unsaturated quantization output and exact consensus. \cite{ma2023novel} also proposed a homomorphic encryption-based cryptographic protocol for energy trading in microgrids. Despite several advantages associated with homomorphic encryption, it has been recently proven to be vulnerable to side-channel attacks \cite{aydin2022exposing}. Wang \textit{et al.} \cite{wang2022efficient} proposed a strategy to achieve privacy-preserved communication in the presence of unreliable insiders and key leakages. The strategy utilizes an evolving-key symmetric encryption-based scheme for ensuring forward secrecy, confidentiality, and non-repudiation in the communication network associated with microgrid controllers. Despite several advancements aimed at enhancing classical cryptographic protocols, the introduction of quantum algorithms into the mainstream computing world has revealed a host of new vulnerabilities.
For example, while AES is currently considered secure against classical computers, Grover's search algorithm implemented on a high-power quantum computer, can break its encryption by rapidly reducing the number of operations required to brute-force cryptographic keys \cite{grassl2016applying}.
Recognizing this threat, several researchers have proposed the use of quantum cryptography for secure communication in cyber-physical microgrids \cite{yan2022quantum, tang2020quantum, tang2020}. Quantum Key Distribution (QKD)
can address the issue of key leakages by utilizing quantum mechanics principles to ensure the secure exchange of secret keys between parties.

A simulation framework was proposed in \cite{tang2020quantum} to simulate QKD protocols in a Real-Time Digital Simulator (RTDS)-based microgrid environment. Lardier \textit{et al.} \cite{lardier2019quantum} designed an open-source co-simulator framework that facilitated the integration of quantum communication with electric grid topologies. Yan \textit{et al.} \cite{yan2022quantum} proposed a QKD-based distributed control framework for microgrid security enhancement. To achieve a higher degree of practicality, the strategy utilized a measurement-device-independent QKD-based setup to defend against side-channel attacks and ensure secure data transmission. The authors also achieved real-time parameter tuning via deep neural networks. Tang \textit{et al.} \cite{tang2022enabling} proposed the use of Software-Defined Networks  (SDNs) for achieving a higher degree of resiliency in practical quantum microgrids. Recognizing that QKD-enabled microgrids are vulnerable to Denial-of-Service (DoS) attacks, an SDN-integrated programmable quantum microgrid setup was proposed in \cite{tang2020programmable}, that mitigated such attacks via programmable post-processing of keys. Apart from DoS attacks, another major issue in QKD-enabled microgrid networks is the presence of unauthorized agents attempting to eavesdrop to steal sensitive information. Such unauthorized eavesdroppers in microgrid networks can cause the noise levels in Quantum channels to track an abnormally high trajectory \cite{amer2021introduction}. This is because, unlike classical channels, noise levels in quantum channels are directly related to malicious attempts at gaining information \cite{amer2021introduction}. This can lead to inaccurate agreements regarding the understanding of quantum keys between a transmitting node and the intended recipient leading to erroneous decryption of measurement signals flowing into the microgrid control plane, which can in turn force local secondary controllers to provide erroneous control decisions jeopardizing the stability of the system. Even though this is a highly concerning threat, prior literature on quantum microgrids has not explicitly attempted to address the correlation between microgrid dynamics and the level of attempted eavesdropping on quantum channels.

Addressing the limitations observed, this paper proposes a novel QKD protocol that identifies erroneous signal decryptions by 
observing microgrid properties. This facilitates low-latency key distribution to support real-time microgrid control.
Further, it enables the reconstruction of lost measurement signals via a dynamic adjacency matrix formulation strategy. The major contributions of the paper are listed below:
\begin{itemize}
    \item We propose a novel QKD protocol for secure microgrid communications. This protocol includes conventional QKD authentication features but does not require time-intensive error detection/reconciliation via classical communication channels. This is a major advantage as time-intensive computations may destabilize the microgrid.
    \item We analyze the impact of node-level measurement manipulations and quantum errors caused by eavesdroppers on conventional QKD-microgrid networks and design a distributed quantum bit error rate (QBER) detection strategy that analyzes observable secondary control signals to identify them. 
    \item Finally, we create a dynamic adjacency matrix formulation strategy that allows the incorrectly decrypted measurement signal to be reconstructed in a trustworthy, multi-hop manner. This strategy enables the QKD-enabled microgrid network to perform nominal operations in the presence of attempted key measurements and node-level manipulations provided at least one microgrid node can transmit reliable sensor measurements without interruption.
\end{itemize}

The remaining parts of the paper are structured as follows. Section II presents an overview of the background knowledge about privacy risks in cyber-physical microgrids and their possible alleviation via QKD. Section III introduces the generic control structure of AC microgrids and security/privacy issues. A detailed explanation of QKD and associated vulnerabilities is provided in Section IV. Details of the proposed resilient quantum microgrid system are presented in Section V. Several case studies are presented in Section V to test the efficacy of the proposed QKD protocol. Finally, Section VI concludes the paper.

\section{Background Knowledge}
\begin{figure}
    \centering
    \includegraphics[width=\linewidth]{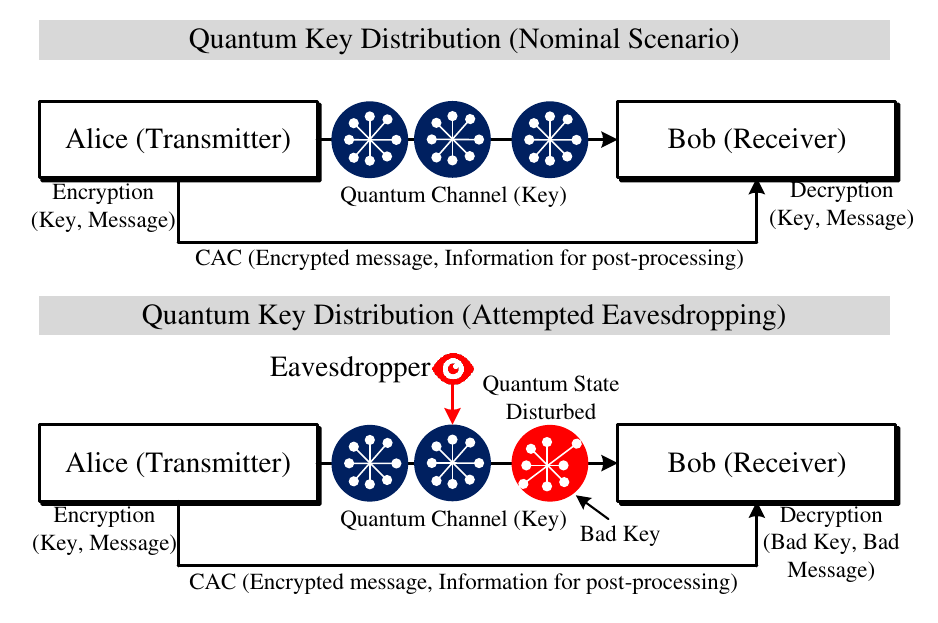}
    \caption{Schematic diagram depicting QKD. CAC is an abbreviation for Classical Authenticated Channel.}
    \label{fig:QKD}
\end{figure}

Microgrid control frameworks are typically distributed in nature to avoid the risk of single-point of failure. This requires that measurement signals need to be sent over a communication network, making them prone to interception and signal tampering \cite{rath2020cyber}. In these settings, attackers may be able to eavesdrop on individual communication channels, compromising the confidentiality of sensitive state information like parameter values and load measurements.
One of the ways to preserve data privacy in the cyber-physical microgrid environment is to use cryptography. However, cryptographic protocols can sometimes be inadequate in fulfilling this role.
For example, if the encryption keys are not managed securely or become compromised, it could lead to unauthorized access, data manipulation, or even complete disruption of the microgrid's operation. While more complex protocols with enhanced security measures exist, the operational diversity of distributed energy resources (DERs) and the multiple communication protocols they rely on makes it challenging to deploy them in microgrids, which is studied in \cite{kirti}.

The scalability and resource constraints (typically imposed by the use of legacy devices) of the DERs within the microgrid environment can limit the feasibility of implementing computationally heavy encryption algorithms. Apart from that, the need for fast, real-time communication and control in microgrids also restricts the use of complex encryption processes, as they could introduce latency and affect system stability.
While generic encryption strategies, such as using strong cryptographic algorithms and implementing robust key management practices, can provide a layer of security for microgrid communication, they may not fully address all vulnerabilities. For example, conventional encryption solutions might not be resilient against quantum computing-based attacks, which have the potential to break traditional cryptographic schemes with significantly increased computational power.

QKD has attempted to address some of these vulnerabilities in microgrid communication \cite{yan2022quantum}. The QKD framework utilizes the concepts of quantum mechanics to securely exchange keys (utilized for encryption/decryption) between two interacting agents \cite{kundu2021mimo,kundu2023mimo}. By utilizing quantum properties like the no-cloning theorem \cite{wootters1982single}, QKD ensures that any attempt to intercept the transmitted key would disturb the quantum state, revealing the presence of an eavesdropper. This framework shows promise as a secure method of key exchange, which could enhance the integrity of microgrid communications. However, while QKD addresses the key distribution problem, it does not eliminate all security risks within the microgrid environment. As shown in Fig. \ref{fig:QKD}, a major risk in QKD-enabled microgrids arises when potential eavesdroppers continue to remain on the network trying to read secret information thereby disrupting the transmitted message and increasing noise level in the Quantum channel \cite{amer2021introduction}. Disruption of the quantum key often leads to incorrect decryption of microgrid measurements affecting the decision-making capabilities of the secondary control layer and creating cascading errors in the primary control signals. Even though a conventional QKD protocol (as shown in Fig. \ref{fig:QKD}) may utilize the classical authenticated channel (CAC) for post-processing including a certain degree of error correction it can also introduce an additional time delay which can destabilize the system dynamics.
\section{Cyber-Physical Microgrids}

\subsection{Control Architecture}
{The microgrid architecture utilized for testing and validation purposes consists of a two-layered hierarchical control framework. The two layers in the framework are discussed in the following text.}

\subsubsection{Primary Control Layer}
\begin{figure}
    \centering
    \includegraphics[width=\linewidth]{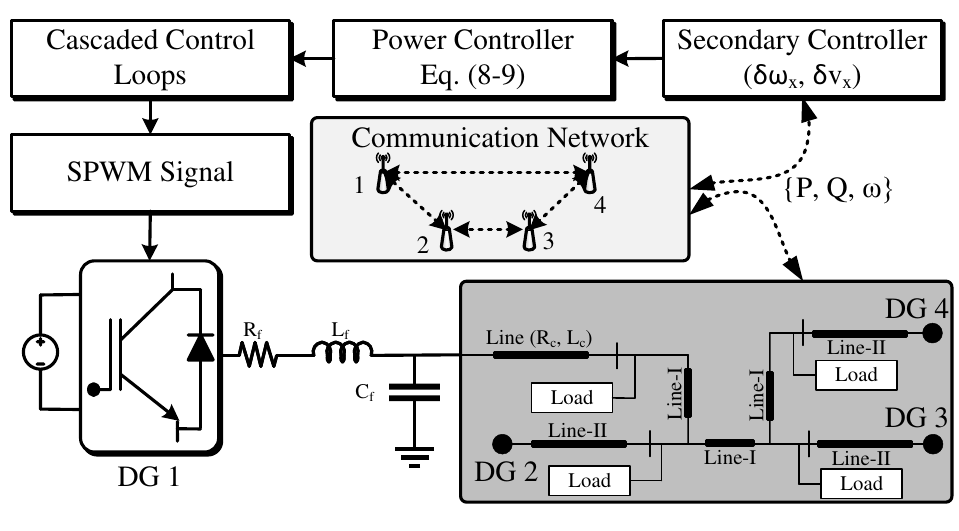}
    \caption{Control architecture of an AC microgrid.}
    \label{fig:mg-control}
\end{figure}
{The primary controller operates as per a droop-based principle \cite{rath2020cyber} to aid in the synchronization of frequency and voltage among distributed energy sources in the microgrid network. This synchronization regulates the power-sharing (both real and reactive power) in the system. The primary power control dynamics can be represented as:}
\begin{equation}
\omega^{\ast}_x = \omega_n-m_{P_x}P_x
\label{prim1}
\end{equation}
\begin{equation}
v^{\ast}_{odx} = v_{odn}-n_{Q_x}Q_x
\label{prim2}
\end{equation}
{where, $\omega^{\ast}_x$ represents system frequency (as measured locally), and $v^{\ast}_{odx}$ represents local voltage. Notations of the form $\omega_{n}$ and $v_{odn}$ represent reference frequency and voltage parameter set-points respectively. $P_x$ and $Q_x$ are locally measured power loads on the $x^{th}$ DG. Droop coefficients are represented by $m_{P_x}$ and $n_{Q_x}$. For a microgrid network with $N$ energy sources, the product of droop coefficients and local power loads are always constant, i.e.,}
\begin{equation}
    m_{P_1}P_1 = m_{P_2}P_2 = ... = m_{P_N}P_N = \Delta\omega_{th}
\end{equation}
\begin{equation}
    n_{Q_1}P_1 = n_{Q_2}Q_2 = ... = n_{Q_N}Q_N = \Delta v_{th}
\end{equation}
{In this context, $\Delta\omega_{th}$ and $\Delta v_{th}$ signify the threshold magnitudes for deviations in frequency and voltage, respectively. Figure \ref{fig:mg-control} illustrates the control architecture of the microgrid as outlined herein.}
{An in-depth discussion of the cascaded outer voltage and inner current control loops can be obtained from \cite{rath2020cyber}.}
{Equations (\ref{prim1}) and (\ref{prim2}) contain negative terms that cause system parameters to incorporate a droop in their normal trajectory. This necessitates parameter restoration that is achieved via the secondary control layer.}

\subsubsection{Secondary Control Layer}
{The microgrid secondary control layer operates on the principle of leader-follower coordination to achieve parameter synchronization between a reference DG (with access to parameter set points) and the rest of the microgrid \cite{rath2020cyber}. This layer is coupled with a distributed communication network to facilitate the exchange of feedback signals among DGs. The objectives of the secondary controller can be described as follows:}
\begin{equation}
\lim_{t \to \infty}||\omega_x(t)-\omega_n|| = 0 \;\forall\; x
\label{ob1}
\end{equation}
\begin{equation}
\lim_{t \to \infty}||m_{Px}{P_x}-m_{Py}{P_y}|| = 0 \;\forall\; x,\;y
\label{ob2}
\end{equation}
\begin{equation}
\lim_{t \to \infty}||n_{Qx}{Q_x}-n_{Qy}{Q_y}|| = 0 \;\forall\; x,\;y
\label{ob3}
\end{equation}
{To accommodate inputs from the secondary layer ($\delta\omega$ and $\delta v$), the power controller equations be reformulated as:}
\begin{equation}
\omega^{\ast}_x = \omega_n-m_{P_x}P_x+\delta \omega_x
\label{sys1}
\end{equation}
\begin{equation}
v^{\ast}_{odx} = v_{odn}-n_{Q_x}Q_x+\delta v_x
\label{sys2}
\end{equation}
{Inputs $\delta\omega_x$ and $\delta v_x$ are computed via single integrator dynamics which are formulated as follows:}
\[
\delta\dot{\omega_x} = K_1\Big(\sum_{l\epsilon{N(k)}}{s_{xy}}(\omega_{l}-\omega_x)+g_x(\omega_n-\omega_x)+
\]
\begin{equation}
\sum_{j\epsilon{N(k)}}{s_{xy}}(m_{P_y}P_{y}-m_{P_x}P_{x})\Big)
\label{sec1}
\end{equation}

\begin{equation}
\delta\dot{v_x} = K_2\Big(\sum_{l\epsilon{N(k)}}{s_{xy}}(n_{Q_y}Q_{y}-n_{Q_y}Q_{y})\Big)
\label{sec2}
\end{equation}
{In this scenario, $s_{xy} \in S_i$ identifies an element within the adjacency matrix $S_i$, corresponding to the bidirectional microgrid communication network graph. This graph comprises pairs of one-way communication links and highlights the condition (either active or inactive) of the connection responsible for communicating the set of elements $\{\omega_x, P_x, Q_x\}$ from the $x^{th}$ DG to the $y^{th}$ DG. The term $g \in G$ denotes the pinning gain value, while $K_1$ and $K_2$ are constants, the determination of which is based on the equation provided below:}
\begin{equation}
K_1 = K_2 \geq\frac{1}{2\lambda_{min}(L+G)} 
\end{equation}
{Here, $L$ denotes the Laplacian matrix. $\lambda_{min}$ represents the smallest eigenvalue for matrix $(L + G)$.}

\textit{{Theorem 1:}}
{The modified power control equations as depicted in equations (\ref{sys1}) and (\ref{sys2}) can achieve the target objectives depicted in equations (\ref{ob1})-(\ref{ob3}) by using inputs $\delta \omega$ and $\delta v$, if the communication network (graph) forms a connected and balanced spanning tree with at least one vertex to accessing parameter references for set-point tracking \cite{rath2020cyber}.}

{\textit{{{Proof:}}} The theorem stated above can be proved if the system in equation (\ref{sys1}) is found to be stable when the communication network is a fully connected and balanced spanning tree. We use Lyapunov Stability Analysis to establish the stability of the depicted system. Differentiating (\ref{sys1}) \textit{w.r.t.} time yields,}
\begin{equation}
\dot{\omega}^{\ast}_x = -m_{P_x}\dot{P}_x+\dot{\delta \omega_x}
\label{th1}
\end{equation}
{Consider variable $y$ such that,}
\begin{equation}
y_x = \omega^{\ast}_x+m_{P_x}P_x
\label{th2}
\end{equation}
{From equation (\ref{sec1}), equation (\ref{th1}) may be reformulated as}
\begin{equation}
\dot{y}_x = \sum_{l\epsilon{N(k)}}{K_1s_{xy}}(y_{l}-y_x) - K_1g_xy_x + K_1g_x(\omega_n+m_{P_x}P_x)
\label{th3}
\end{equation}
{We select a positive, definite Lyapunov Function Candidate (LFC) as depicted below to demonstrate stability.}
\begin{equation}
V = \frac{1}{2}\sum_{k=1}^{N}y_x^2
\label{th4}
\end{equation}
{Using equations (\ref{th3}) and (\ref{th4}), $\dot V$ can be derived as:}
\[
\dot{V} = \sum_{k=1}^{N}\sum_{l\epsilon{N(k)}}{K_1s_{xy}y_x(y_x-y_y)}
\]
\begin{equation}
-\sum_{k=1}^{N}\Big(K_1g_xy_x^2-K_1g_xy_x(\omega_n+m_{p_x}P_x)\Big)
\label{th5}
\end{equation}
{The following condition always holds as the network is a spanning tree.}
\begin{equation}
\sum_{k=1}^{N}\sum_{l\epsilon{N(k)}}{s_{xy}y_x^2} = \sum_{k=1}^{N}\sum_{l\epsilon{N(k)}}{s_{xy}y_y^2}
\label{th6}
\end{equation}
{From equation (\ref{th5}) and equation (\ref{th6}), $\dot V$ can be reformulated as:}
\[
\dot{V} = -\frac{1}{2}\sum_{k=1}^{N}\sum_{l\epsilon{N(k)}}{K_1s_{xy}(y_x-y_y)^2}
\]
\begin{equation}
-\sum_{k=1}^{N}{K_1g_xy_x^2} + \sum_{k=1}^{N}{K_1g_xy_x(\omega_n + m_{P_x}P_x)}
\label{th7}
\end{equation}
{$\dot V$ can be bounded as:}
\[
\dot{V}\leq -\frac{1}{2}\sum_{k=1}^{N}\sum_{l\epsilon{N(k)}}{K_1s_{xy}(y_x-y_y)^2}
\]
\begin{equation}
-\sum_{k=1}^{N}{K_1g_xy_x^2} + \sum_{k=1}^{N}{K_1g_x(\omega_n + m_{P_x}P_x)^2}
\label{th8}
\end{equation}
{As $y_x^2 \geq (\omega_n + m_{Px}P_x)^2 \Rightarrow \dot V \leq 0$ $\Rightarrow$ system depicted in (\ref{sys1}) is stable $\Rightarrow$ (\ref{th3}) shall converge to zero at steady state, achieving the objectives depicted in (\ref{ob1}) and (\ref{ob2}). The same approach can be followed to prove that the system depicted in (\ref{sys2}) will achieve the objective in (\ref{ob3}) if the network forms a spanning tree. This proves the validity of Theorem 1.}

{Note that for any given graph, a spanning tree is not unique \cite{rath2020cyber}. This means that we can create a set of spanning trees defined by adjacency matrices $S = \{S_1, S_2, ...., S_n\}$ all of which can be utilized to achieve the control objectives in equations (\ref{ob1})-(\ref{ob3}). Another noteworthy point is that all possible directed edges in the communication graph belong to at least one adjacency matrix in $S$.}

\subsection{Attack Modeling}
\begin{figure}
    \centering
    \includegraphics[width=0.45\linewidth,clip,trim={6 6 6 137}]{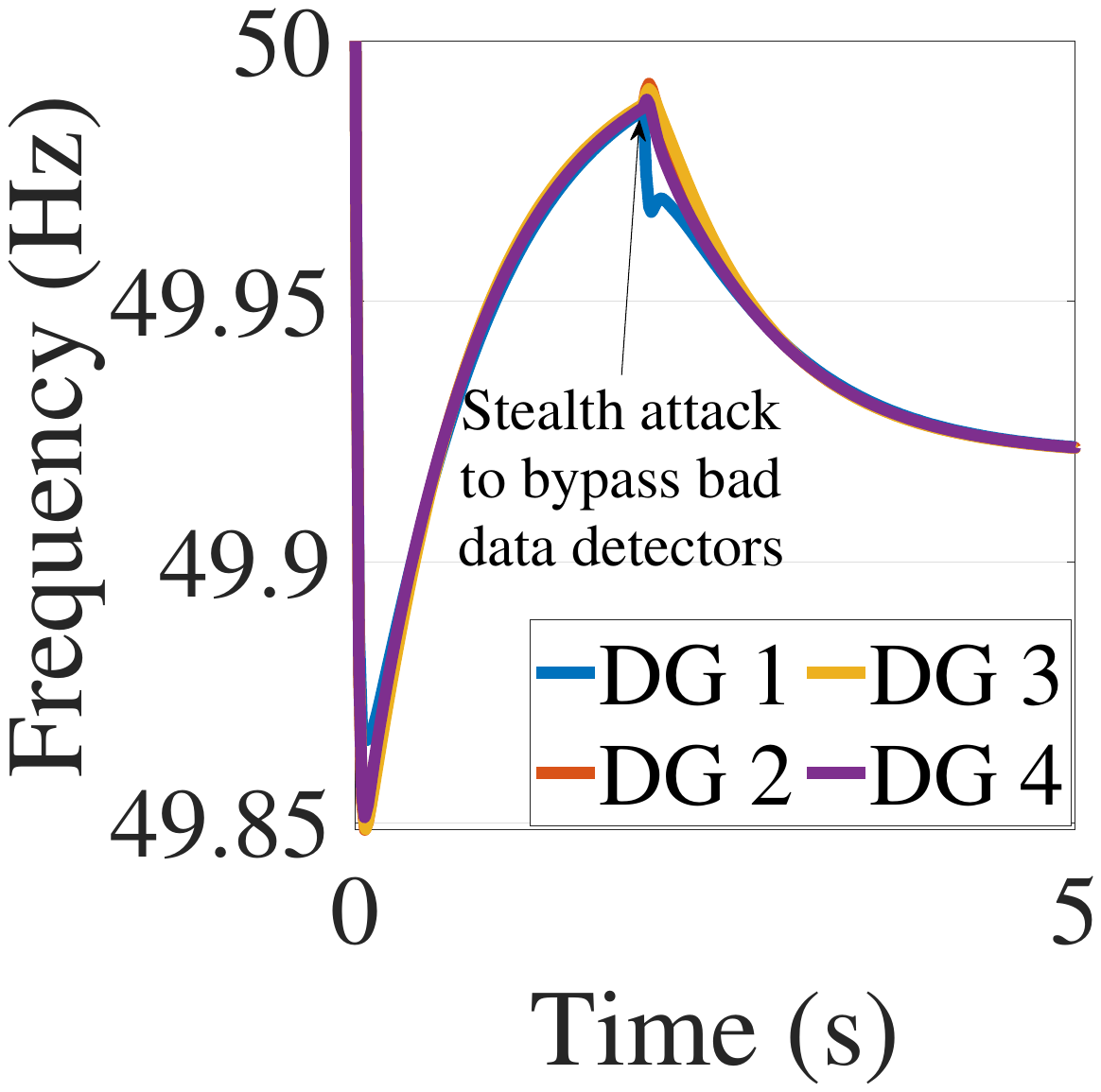}
    \includegraphics[width=0.45\linewidth,clip,trim={6 6 6 121}]{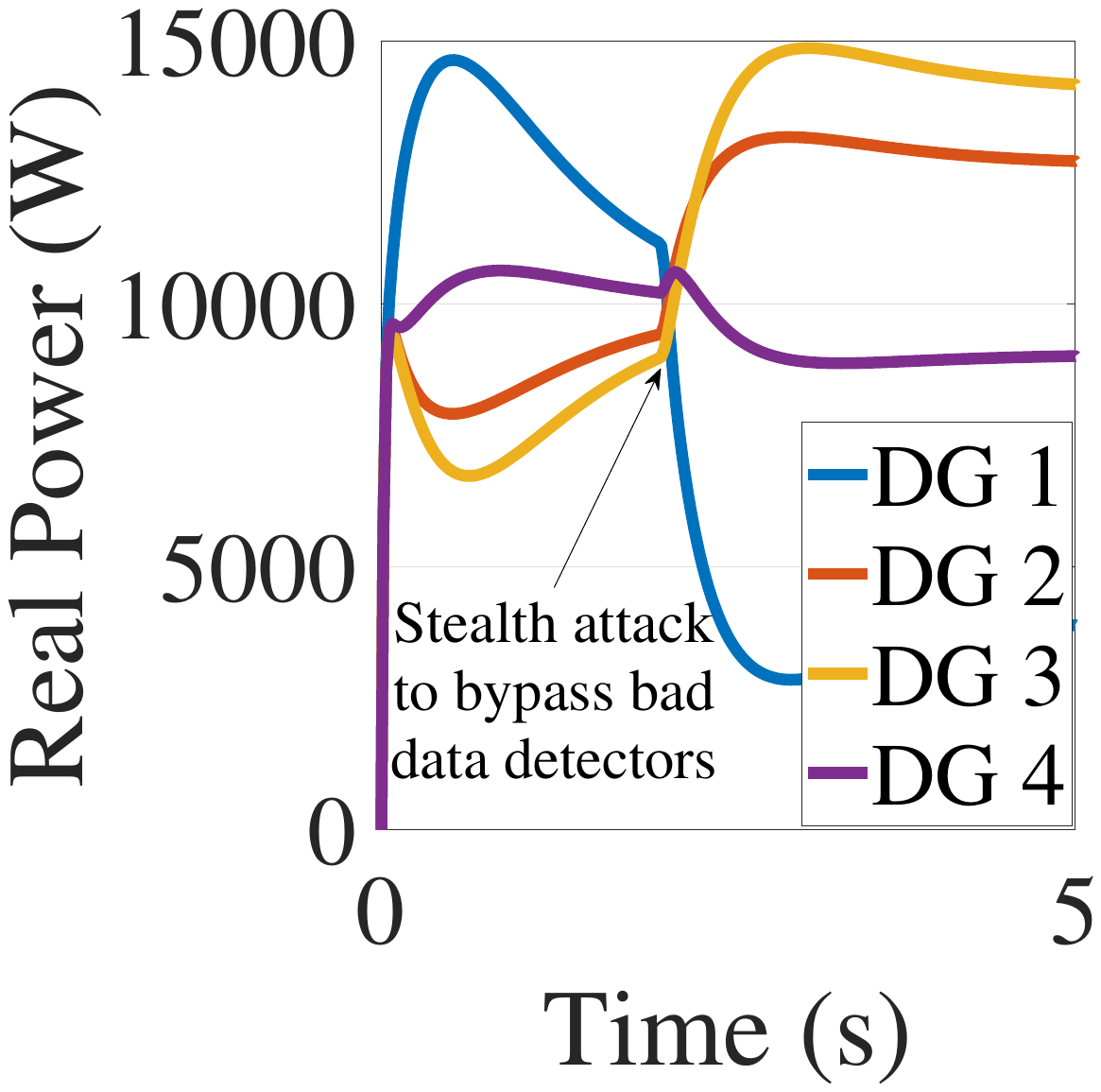}\\[-4ex]
    \caption{{Stealth attacks targeting system frequency and real power sharing designed by attackers who could eavesdrop on microgrid state information flowing via the communication layer.}}
    \label{fig:no_quantum}
\end{figure}

{An attacker targeting the microgrid can have control over certain DGs and/or network connections linked to at least one adjacency matrix from $S_a \subset S$ (used to achieve system convergence). This ability indicates that the attacker has exploited internal privileges or external breaches to introduce malicious software into on-site computing devices and/or sensor units. When extended to communication devices, the attack surface can allow the injection of malicious data into legitimate data flows. A careful attacker typically seeks to achieve a disguised manipulation, attempting to mask false data to mislead operators who are monitoring the system. In the event of a malware-induced cyber-attack, the system controller is altered to reflect a modified signal $c^{alt}$, as per the following formula:}
\begin{equation}
    {{c^{alt}(t)} = f({x_n(t)}) + {\Xi_A(t) \cdot x_a(t)}}
    \label{attack}
\end{equation}
{Here, $x_n(t)$ denotes the normal state vector at time step $t$, while $x_a$ represents the vector introduced by the attacker, and $f(\circ)$ symbolizes the function that maps the state vector to the control vector under normal circumstances. $\Xi_A(t)$ is defined as an adjacency matrix that signifies the connections through which tainted (manipulated) data is disseminated.
As a result of the attack vector, selecting ${S_i} \in S_a$ to serve as the operational communication structure will result in a deviation from the expected steady-state path.}
{A major requirement for attackers to execute an undetectable stealth attack is that they should be able to eavesdrop on the communication channels associated with the microgrid network and learn system information \cite{rath2022behind}. For example, an eavesdropper that can observe microgrid state parameter values in real time can establish a set of upper and lower bounds for its false data injections to bypass scrutiny from bad data detectors. This results in a bounded stealth attack that can manipulate system stability as shown in Fig. \ref{fig:no_quantum}. Eavesdropping is generally possible if the communication network uses a weak cryptographic protocol.}


\section{Quantum Cryptography for Security of Microgrids}
QKD can address the cybersecurity risks presented in the preceding section by enhancing the level of undetectability associated with cryptographic keys used in the microgrid environment. This section presents our proposed quantum cryptographic scheme for secure communication in the cyber-physical microgrid environment as depicted in section III. The proposed scheme involves the use of an optical fiber network and a classical communication network connecting all the DGs in the network. The classical network is responsible for facilitating the exchange of symmetrically encrypted local measurement signals $\{\omega_x(t), P_x(t), Q_x(t)\}$ between the $x^{th}$ DG and its neighbors for the computation of local secondary control signals. The keys for decrypting the encrypted measurement signals are transmitted via the network of optical fibers (quantum channel). Acknowledging the possibility of events where the exchanged keys are erroneous (potentially leading to incorrect decryption and subsequent instability), we utilize a DG-level localized scheme to identify and reconstruct incorrectly decrypted signals.

\subsection{QKD-integrated System Model}
The proposed QKD-integrated microgrid model is a fully connected graph consisting of $2M$ information exchange links (including $M$ classical communication channels and $M$ optical fibers) and $N$ DGs (nodes). All classical communication in the system is encrypted via symmetric cryptography. The keys required for the decryption of these messages are exchanged through the optical fiber network. This network is essential for transmitting individual quantum states, primarily photons, that have been polarized to represent quantum bits or qubits. The transmitting DG prepares these qubits by selecting their polarization states, such as horizontal or vertical, and the (intended) recipient DG, measures the polarization of the received photons. In our system, the transmitter node prepares a series of qubits in random states from one of two bases: rectilinear \( \{|0\rangle,|1\rangle\} \) or diagonal \( \{|+\rangle,|-\rangle\} \). The states \( |+\rangle \) and \( |-\rangle \) are defined as:
\begin{equation}
     |+\rangle =  \frac{|0\rangle + |1\rangle}{\sqrt{2}} 
\end{equation}
\begin{equation}
    |-\rangle = \frac{|0\rangle - |1\rangle}{\sqrt{2}}
\end{equation}
Furthermore, over the CAC (as depicted in Fig. \ref{fig:QKD}), the transmitting node informs the recipient node of the bases used for each qubit, but not the actual states. The intended recipient then discloses which of its measurements used the correct basis. They discard all bits corresponding to measurements associated with the wrong basis.
The recipient randomly chooses one of the two bases to measure each qubit and records the result and the basis used for each measurement.
The use of the QKD in addition to the generic microgrid communication setup ensures that any intercepted encrypted message remains incomprehensible without the shared quantum key.

To ensure a secure and reliable state update in the QKD-based microgrid environment, it's crucial to ensure the integrity of quantum keys, the authenticity of physical agent measurements, and the conformity of node operations with predefined system properties. With QKD, eavesdropping or data tampering risks are minimized, but several challenges remain. Potential threats include adversaries attempting to breach system security, for example, by incorporating malicious manipulations in nodal measurement signals before encryption. Since QKD operation is restricted to the communication plane, a DG-level, local authentication mechanism becomes essential to identify and counteract signals manipulated before encryption.
Apart from node-level attacks, QKD can also be vulnerable to eavesdropping attempts. Malicious entities can attempt to read quantum states, which even though theoretically not possible (due to the no-cloning theorem \cite{wootters1982single} and observer effect \cite{sassoli2013observer}), can lead to disturbance of the original quantum state thereby increasing the noise magnitude \cite{amer2021introduction}. This has a proportional effect on the error rate associated with quantum keys.
In traditional quantum networks, the transmitting and the receiving DG publicly compare a random subset of their remaining bits to estimate the QBER. If the error rate is above a certain threshold, they discard the key; otherwise, they use error correction and privacy amplification to generate a secure key. QBER is formulated as the number of discrepancies ($e_D$) in the shared sample to the total number of bits ($\zeta$) in the sample i.e.,
\begin{equation}
    \text{QBER} = \frac{e_D}{\zeta}
    \label{QBER}
\end{equation}
A non-zero QBER indicates the presence of errors. In the context of microgrids, dropping a key and subsequent re-transmission may lead to additional computation/communication latency and incorrect decryption of real-time measurement data packets, creating instability in the network. Hence, it is essential to quickly reconstruct signals that can be otherwise lost. To achieve this, we use a microgrid-system-informed mechanism for real-time detection and reconstruction of measurement signals incorrectly decrypted due to high QBER values. This mechanism also allows us to sense the presence of QBER values using inherent microgrid physics without requiring additional computation as depicted in equation (\ref{QBER}).

\subsection{Error Detection}
\begin{table}
	\renewcommand{\arraystretch}{1.3}
	\caption{Event-Triggered Error Detection Metrics \cite{sahoo2018stealth, sahoo2019detection}}
	\label{tab:2}
	\centering
\begin{threeparttable}
	\begin{tabular}{c|c|c}
		\hline
		Affected Signal & Detection Criteria & Terminology \\
		\hline
		\hline
		
		$\omega_y, P_y$ & \begin{tabular}{@{}c@{}} $|g_x(\omega_n - \omega_x) + \sum{s_{xy}}[(\omega_y-\omega_x)$\\$+ (m_{P_y}P_{y}-m_{P_x}P_{k})]|> \Upsilon_1$ \end{tabular}& $DM^k_1$ \\
		\hline
		
		$v_y, Q_y$ & \begin{tabular}{@{}c@{}} $|\sum {s_{xy}}(n_{Q_y}Q_{y}-n_{Q_y}Q_{y})| >$  $\Upsilon_2$ \end{tabular} & $DM^k_2$\\
		\hline
	\end{tabular}
\end{threeparttable}
\end{table}
The usage of erroneous keys in the traditional Quantum-enabled microgrid network can lead to uncontrollable deviations in the state trajectory, causing possible instabilities. As a consequence of the \textit{Observer effect} \cite{sassoli2013observer}, erroneous keys may be generated due to the presence of eavesdroppers who are attempting to measure Quantum states. Additionally, the typical QKD network cannot protect against cyber-attacks that are executed via malware lodged within microgrid DG-level cyber-physical devices. Malware lodged within DG-level devices can introduce false signals into the classical communication network and masquerade them as \texttt{True} measurements via fake quantum keys. As these vulnerabilities were formulated to prey on generic quantum microgrids, we propose a detection framework that leverages microgrid system properties to identify erroneously decrypted measurement values or false inputs fed into the quantum network by malware-infected internet-of-things (IoT) devices. The proposed distributed detection framework is designed to identify real-time violations resulting in asynchrony among DGs in the network by monitoring local secondary control signal magnitudes. More specifically, we formulate two detection metrics $DM^k_1$ and  $DM^k_2$, that represent error signals signifying the mismatch between measurements from two or more generators in the network. If an unexpected attack vector that performs manipulations at the node level is encountered, we hypothesize that the magnitude of metric,  $DM^k$ will increase beyond the threshold specified in Table \ref{tab:2}. The same deviation will also be observed when a recipient node in the quantum microgrid performs an incorrect decryption. Hence, the metrics also form an effective triggering framework that alerts the system about the possible presence of eavesdroppers trying to read quantum states and generating high QBER in the process, leading to the activation of a signal reconstruction framework that replaces the affected signal. More details about the proposed signal reconstruction framework are provided below.

\subsection{Signal Reconstruction via Dynamic Adjacency Matrices}
\begin{figure}
    \centering
    \includegraphics[width=\linewidth]{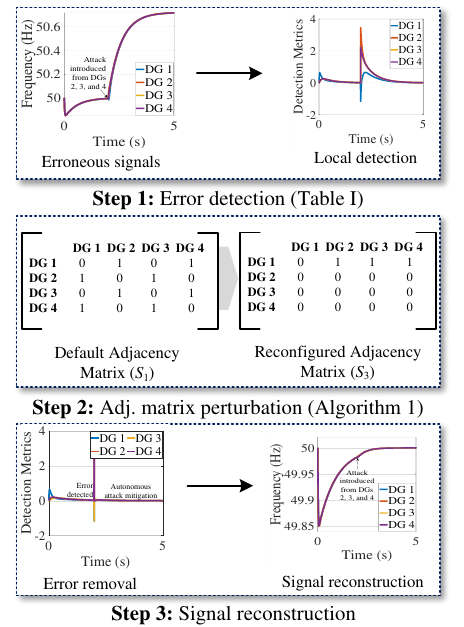}
    \caption{Steps involved in the proposed error detection and mitigation framework.}
    \label{steps}
\end{figure}

As depicted in the preceding section, we can identify the presence of incorrect measurements on the recipient side via the rule-based mismatch detection criteria as depicted in Table \ref{tab:2}. The detection of a high-magnitude mismatch triggers a reconstruction framework that modifies the active communication matrix elements to alter the current secondary control topology. The signal reconstruction mechanism utilized to perform the recovery of incorrectly decrypted measurement signals (or manipulated measurement values) exploits the system's ability to achieve secondary control objectives via any communication adjacency matrix that can establish a spanning tree topology, connecting all DGs \cite{rath2020cyber} in the quantum microgrid network. Let ${S_{i}} \in {S}$ be the set of available adjacency matrices that allow the microgrid's secondary control layer to achieve its intended target objectives. To represent the errors introduced by both eavesdroppers and node-level attackers, we use a generalized form of $\Xi_A$ called the error propagation matrix (represented as $\Xi_E$). In this setup, the system can be manipulated by an adversary if and only if:
\begin{equation}
    {S_{i}} \cap {\Xi_E} \not = 0
\end{equation}
Hence, to establish resiliency in the microgrid network, we simply formulate an adjacency matrix ${S_{j}} \in {S}$ that satisfies the following criterion:
\begin{equation}
    {S_{j}} \cap {\Xi_E} = 0
    \label{mitigation}
\end{equation}
Note that there always exists at least one ${S_{j}}$ that satisfies the criterion formulated above unless all the nodes of the microgrid are attacked simultaneously. This minimalist framework allows us to perform the reconstruction of \texttt{True} signal values to replace all \texttt{False} measurements in the system.
Specific steps involved in the signal reconstruction process following the detection of a \texttt{False} or incorrectly decrypted measurement are depicted in Fig. \ref{steps}.
The steps depicted in the figure ensure that the affected recipient in the quantum-microgrid network is localized, with no inputs from a malicious agent (which may be an infected node or a communication channel linked with an optical fiber from which eavesdroppers are trying to read quantum states). Finally, all manipulated signals are reconstructed and replaced securely. Note that a limitation of this scheme is its inability to reconstruct malicious signals if all elements in the microgrid cyber graph are attacked simultaneously. However, during node manipulations, it can successfully establish resiliency even in the presence of a single agent that is free from attack elements.

\section{Performance Analysis and Simulation Results}
For performance evaluation of the error detection and signal reconstruction mechanism presented in the preceding section, the autonomously operating AC microgrid with $N = 4$ DGs (as depicted in Fig. \ref{fig:mg-control}) is designed in the Simulink environment. Parameter values and topological details for the test system are available in Table \ref{parameters}. The system follows the control architecture depicted in section III. The DGs in this system are fully connected using two parallel layers of communication: (i) a classical network, which is used to exchange encrypted measurements, and (ii) a quantum key distribution network (consisting of optical fibers),  which is used to exchange the keys required for decryption of the messages exchanged via the classical network. We provide two unique case studies to showcase the impact of eavesdroppers who try to measure quantum states in a bid to read the cryptographic keys and formulate the attack vector in equation (\ref{attack}). We also demonstrate the impact of attackers who have DG-level access to directly manipulate measurements even before they are encoded. Finally, we provide several case studies to demonstrate how our proposed detection and mitigation framework provides resiliency against each of the aforementioned examples.

\begin{figure}
    \centering
    \includegraphics[width=\linewidth]{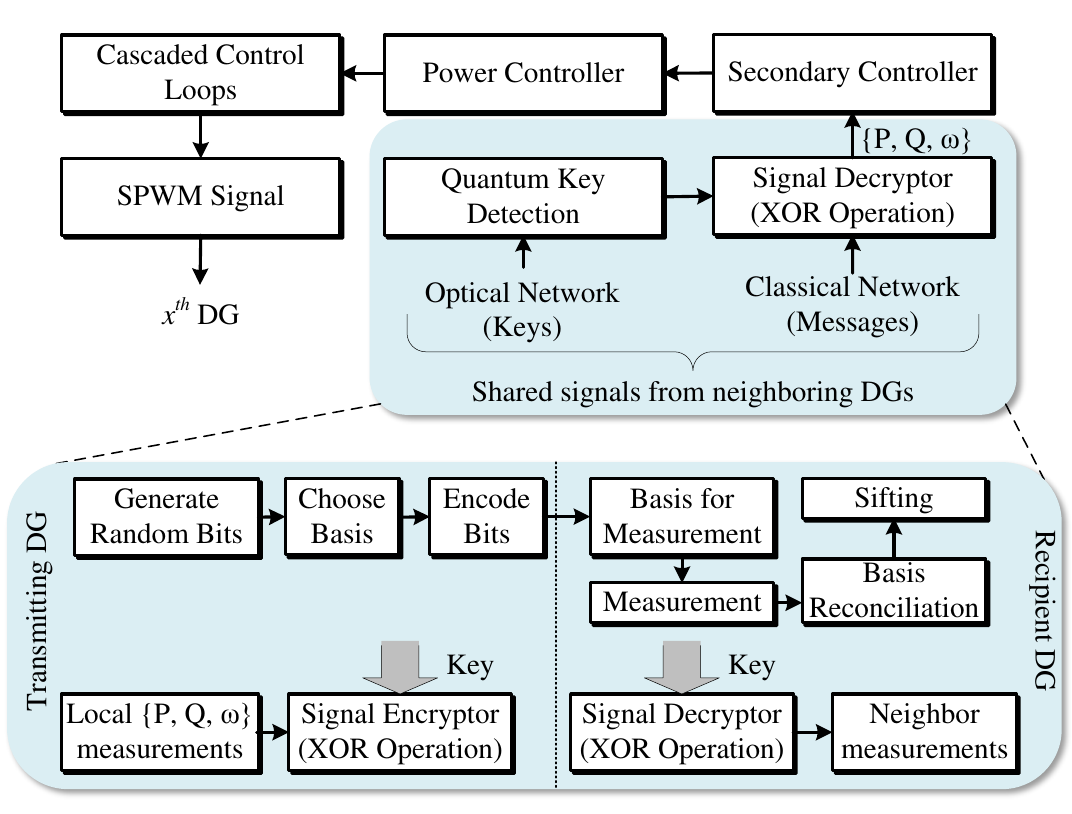}
    \caption{{Simulink-based implementation of local control at the $x^{th}$ DG in a conventional QKD-enabled microgrid network.}}
    \label{fig:block1}
\end{figure}

\begin{algorithm}
\caption{Dynamic Matrix Perturbation}
\label{alg:mat}
\begin{algorithmic}[1]
\State \textbf{Input:} $\{DM_1^k, DM_2^k\}_{k=1}^N$, Adjacency matrix \(S_i\)
\State Create a matrix $S_j$ such that $S_j = S_i$.
\State Initiate $\Xi_E$ as a null matrix.
\State Choose a DG $b$ $\ni$ $(DM_1^b+DM_2^b) = 0$.
\For{each $(DM_1^a+DM_2^a) \not = 0$}
    \For {$\{s_{ak}, s_{ka}\}_{k=1}^N \in S_j$}
    \State $\{s_{ak}, s_{ka}\}_{k=1}^N = 0$
    \State $\{s_{ba}\} = 1$
    \EndFor
    \State Flip $a^{th}$ row elements (i.e., DG $a$) of $\Xi_E$ to 1.
    \State Flip all non-zero diagonal elements of $\Xi_E$ to 0.
\EndFor
\State If any column (except the column corresponding to DG $b$) in $S_j$ is 0, replace its $b^{th}$ element with 1.
\State \textbf{Check:} $S_j \cap \Xi_E = 0$.
\State \textbf{Return:} Perturbed matrix, $S_j$.
\end{algorithmic}
\end{algorithm}

\begin{table}
	\renewcommand{\arraystretch}{1.3}
	\caption{Model Parameter Values}
	\label{tab:3}
	\centering
\begin{threeparttable}
	\begin{tabular}{|c|c|c|c|}
		\hline
		Parameter & Value & Parameter & Value \\
		\hline
		\hline
		
		\(R_f\) & 0.1 \(\Omega\)  & \(C_f\) & 200 \(\mu\)F \\  
		\hline
		\(L_f\) & 4 mH & \(m_P, n_Q\) & \(1\times{10^{-4}}\)  \\ 
		\hline
		\(R_c\) & 0.1 \(\Omega\) & \(w_n\) & \(2\pi50\) rad/s \\ 
		\hline
		 \(f_{sw}\) &  10 kHz &  Line-II & 0.5 mH + 0.07 \(\Omega\) \\ 		
		\hline 
            Line-I & 1.5 mH + 0.1 \(\Omega\) & \(V_{dc}\) & 1000 V \\
            \hline 
	\end{tabular}
\end{threeparttable}
\label{parameters}
\end{table}

\begin{figure}
    \centering
    \includegraphics[width=0.45\linewidth,clip,trim={6 6 6 137}]{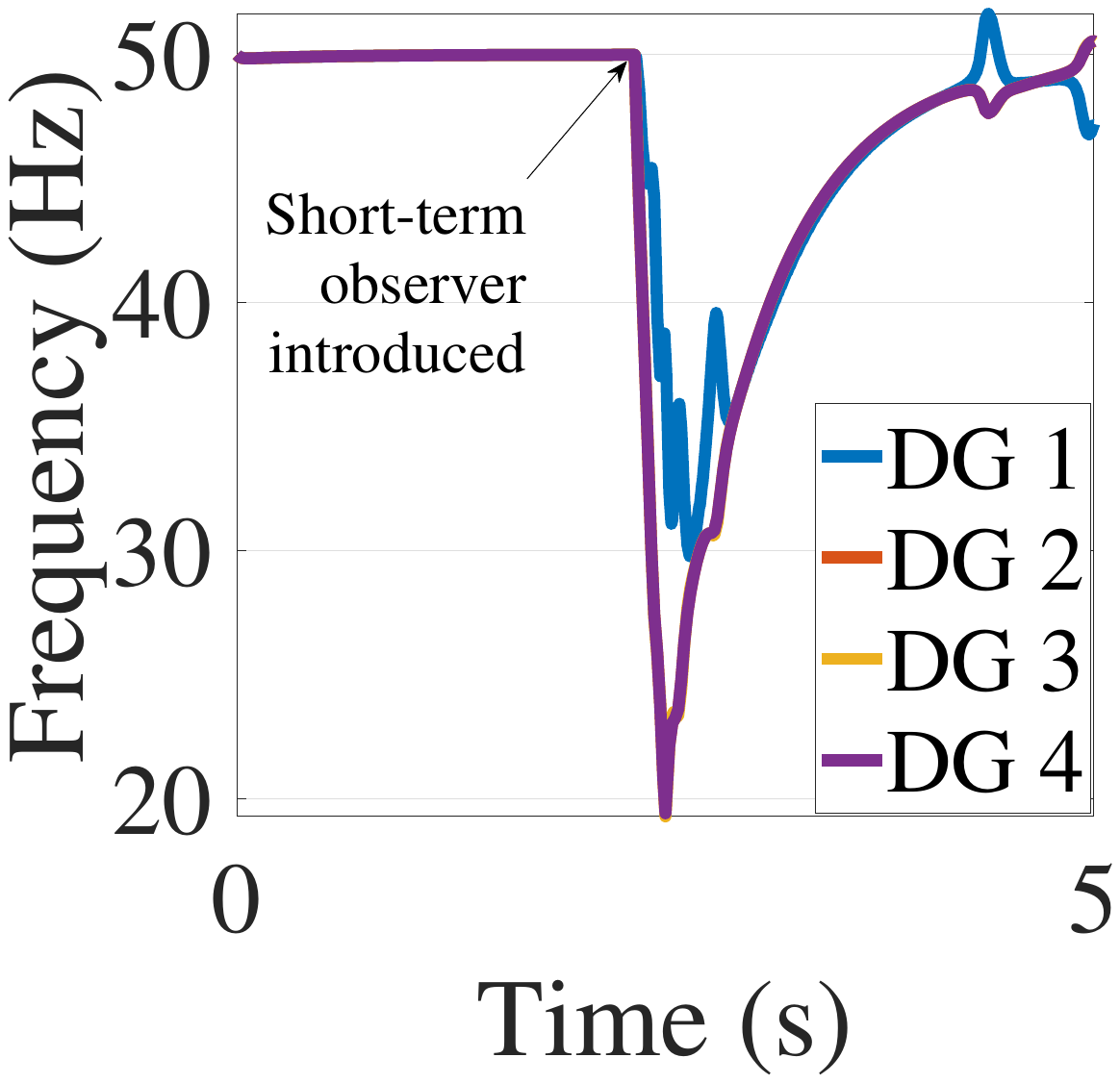}
    \includegraphics[width=0.45\linewidth,clip,trim={6 6 6 121}]{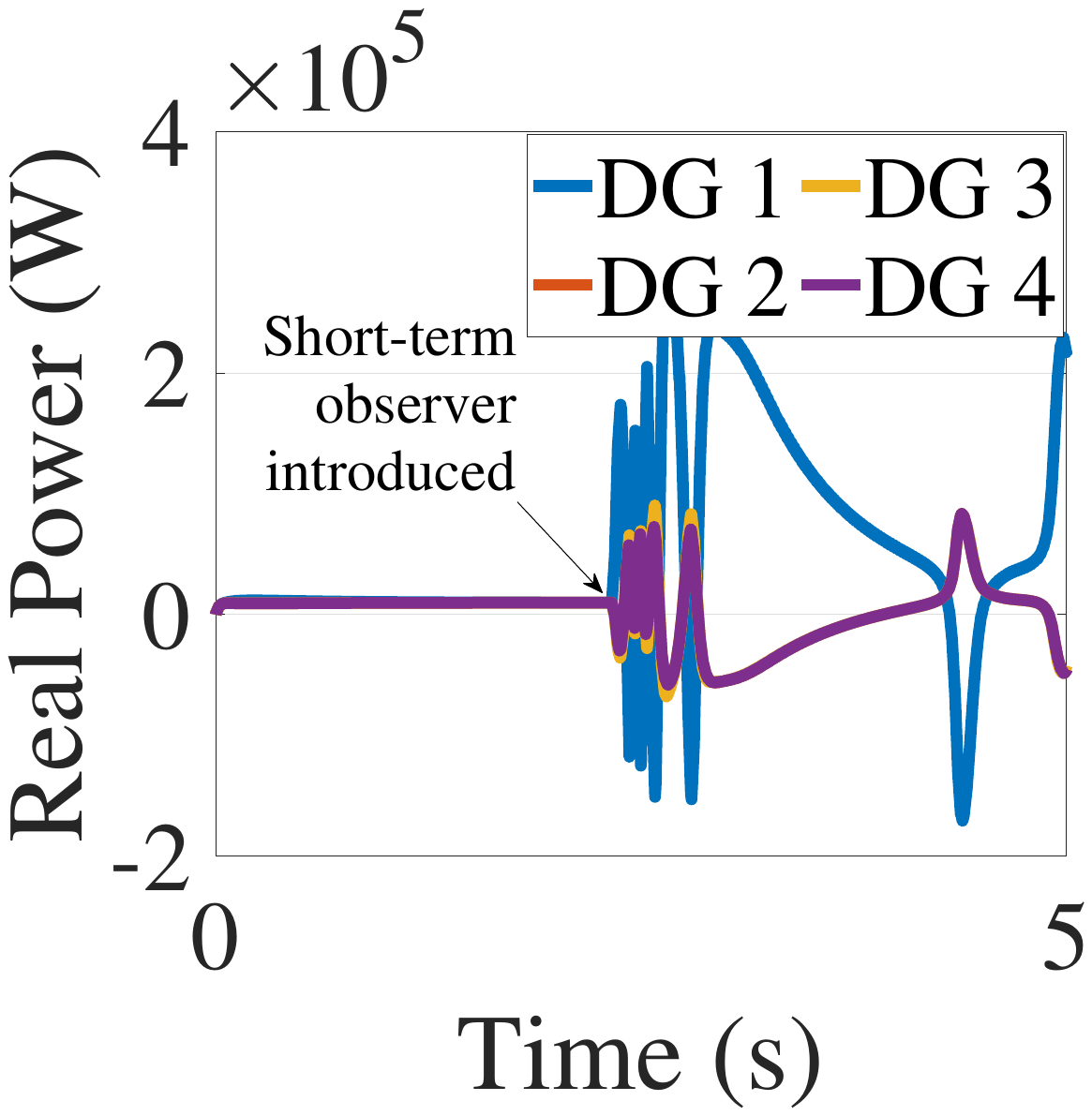}\\[-4ex]
    \includegraphics[width=0.45\linewidth,clip,trim={6 6 6 121}]{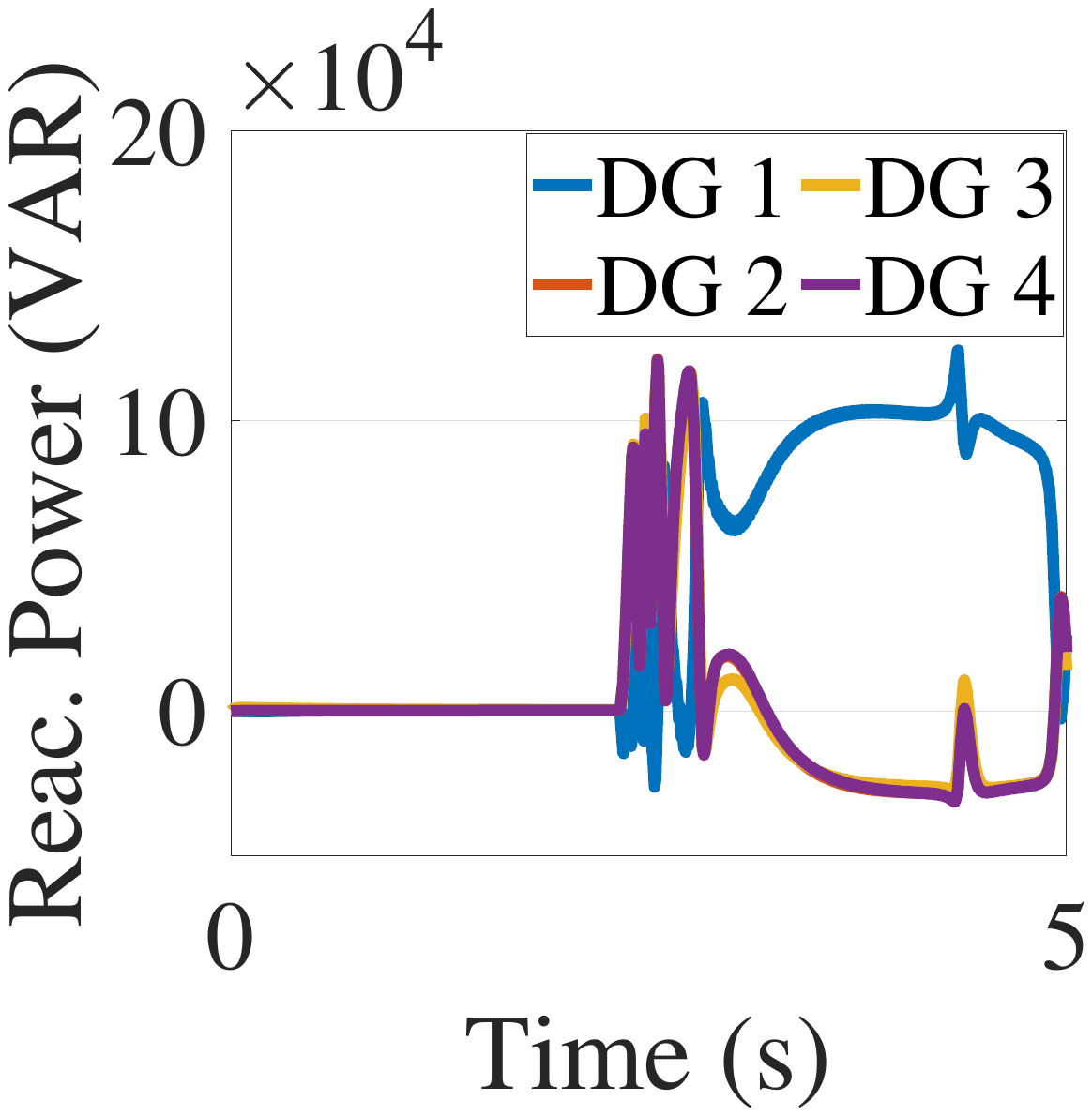}
    \includegraphics[width=0.45\linewidth,clip,trim={6 6 6 137}]{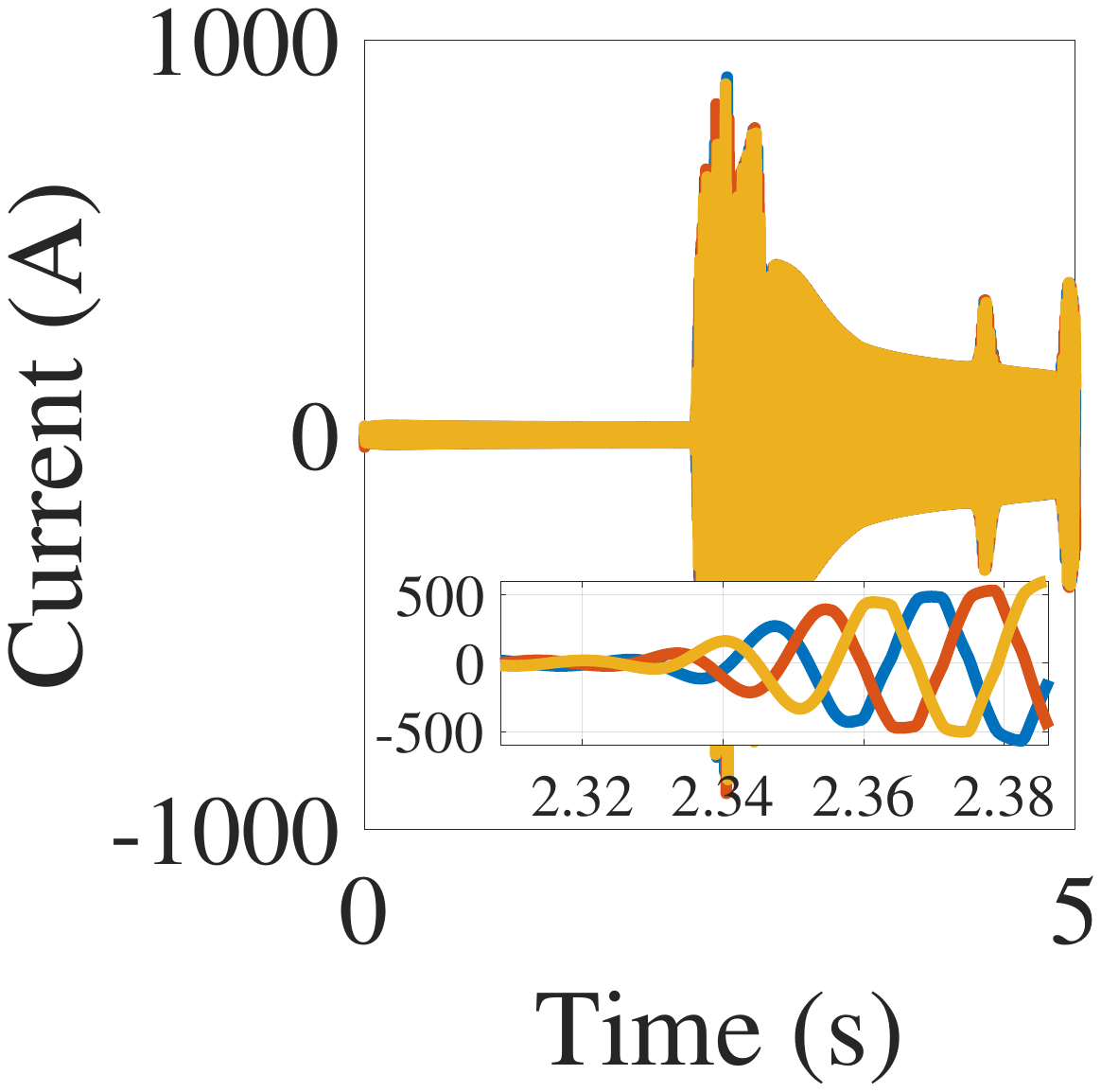}\\[-4ex]
    \includegraphics[width=0.45\linewidth,clip,trim={6 6 6 137}]{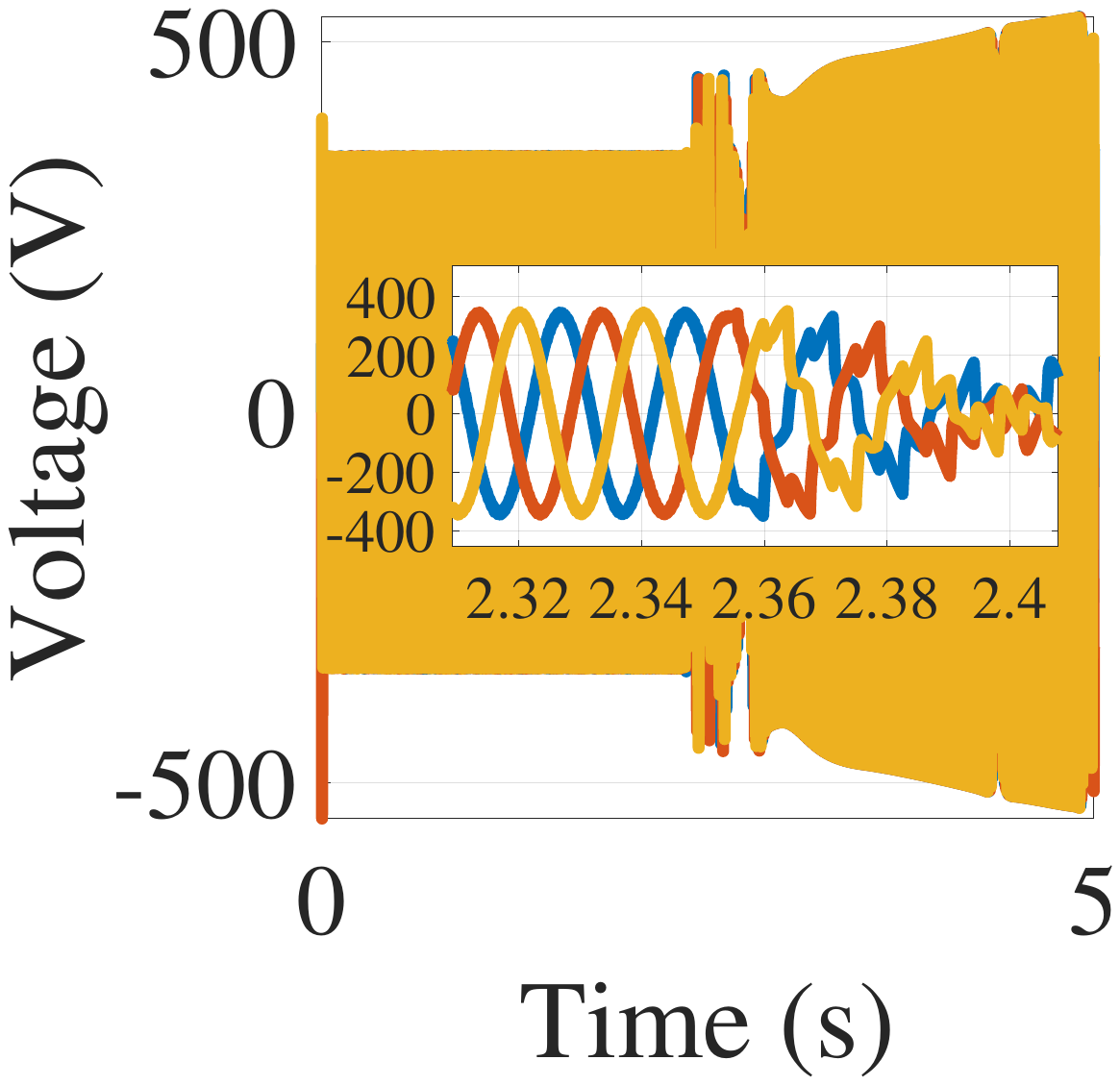}
    \includegraphics[width=0.45\linewidth,clip,trim={6 6 6 121}]{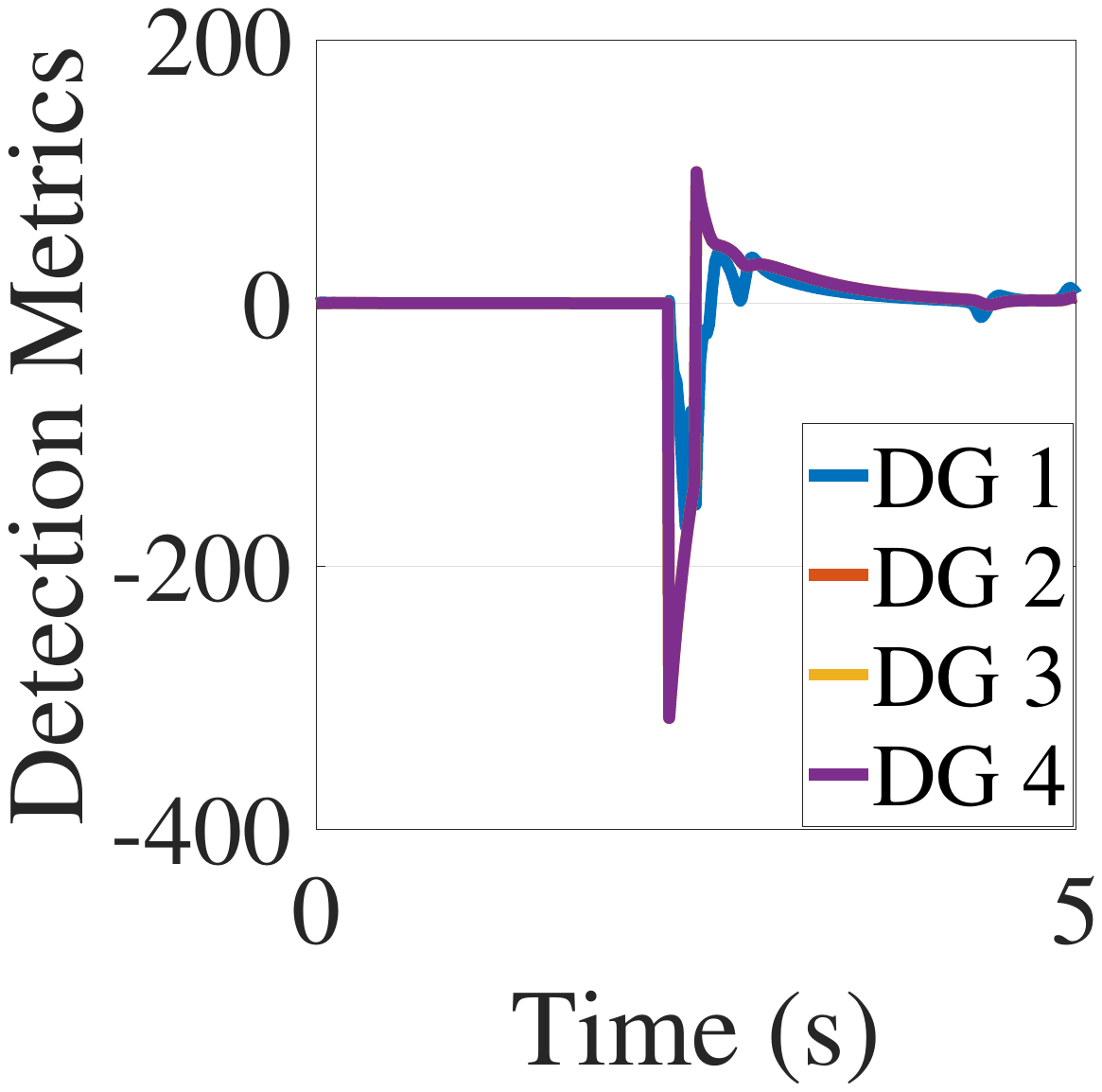}\\[-4ex]
    \caption{{Impact of the presence of a short-term observer who tries to observe quantum states from $t = 2.32 s - 2.5 s$ on a generic QKD-based microgrid.}}
    \label{fig:study1}
\end{figure}

\subsection{Consequences of Erroneous Quantum Keys}
In the following case studies, we utilize the quantum AC microgrid testbed depicted in the preceding paragraph by plugging out the proposed fortification framework. The test system's local DG-level signal processing and control schematics are shown in Fig. \ref{fig:block1}. We subject this testbed to two unique scenarios. In the first scenario, an eavesdropper (the attacker) tries to observe the quantum channel in Fig. \ref{fig:QKD} to learn the cryptographic keys used to encode measurement signals at the current time step. 
%
%
However, upon realizing that an accurate observation is not possible, immediately stops its attempts. This is referred to as 'short-term observation'. In the second case, the observer (also called the eavesdropper or the attacker) chooses to stay within the system and continues to eavesdrop as it is persistently attempting to manipulate the system dynamics. This is called 'persistent observation'. 

\begin{figure}
    \centering
    \includegraphics[width=0.45\linewidth,clip,trim={6 6 6 137}]{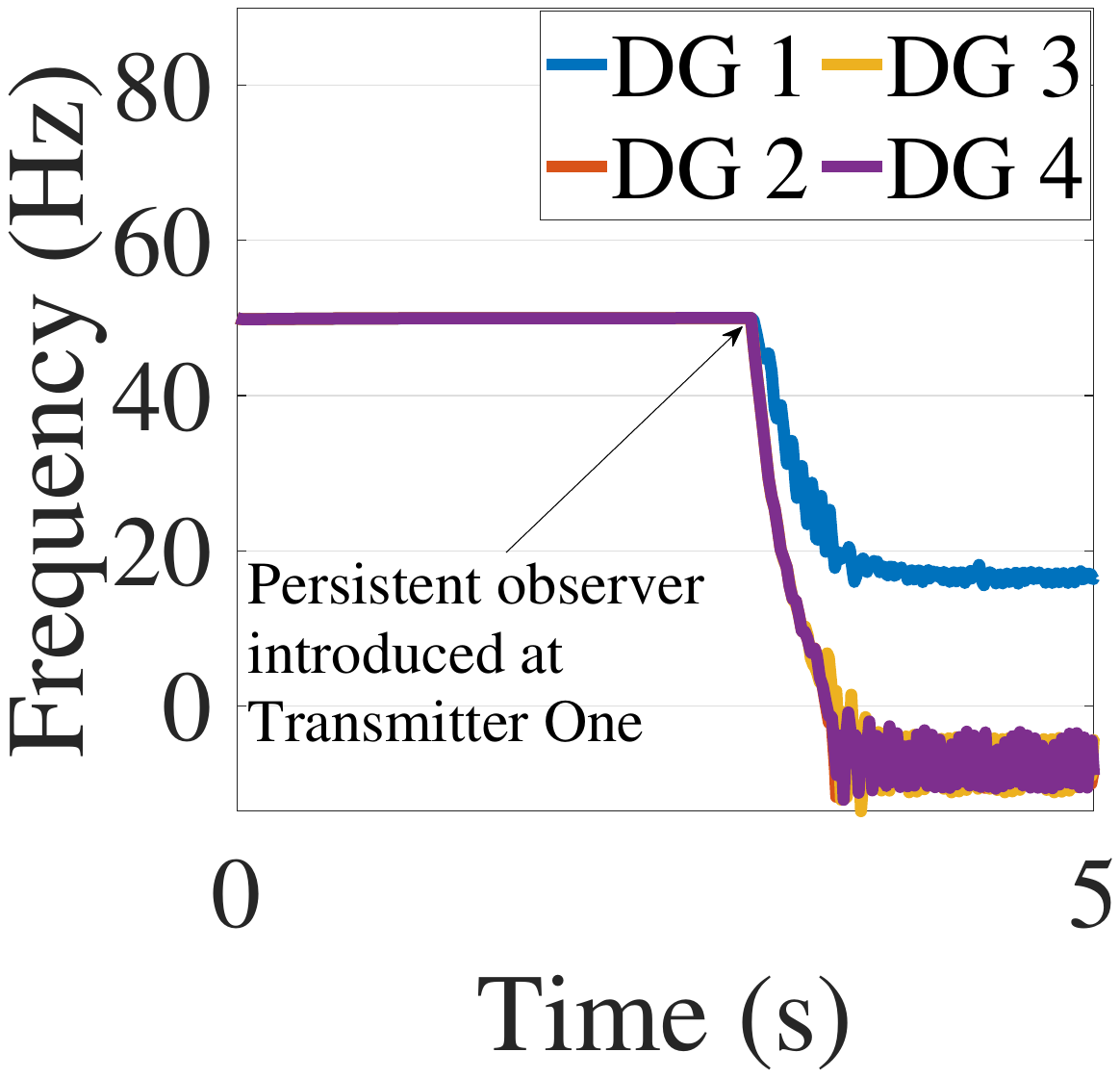}
    \includegraphics[width=0.45\linewidth,clip,trim={6 6 6 121}]{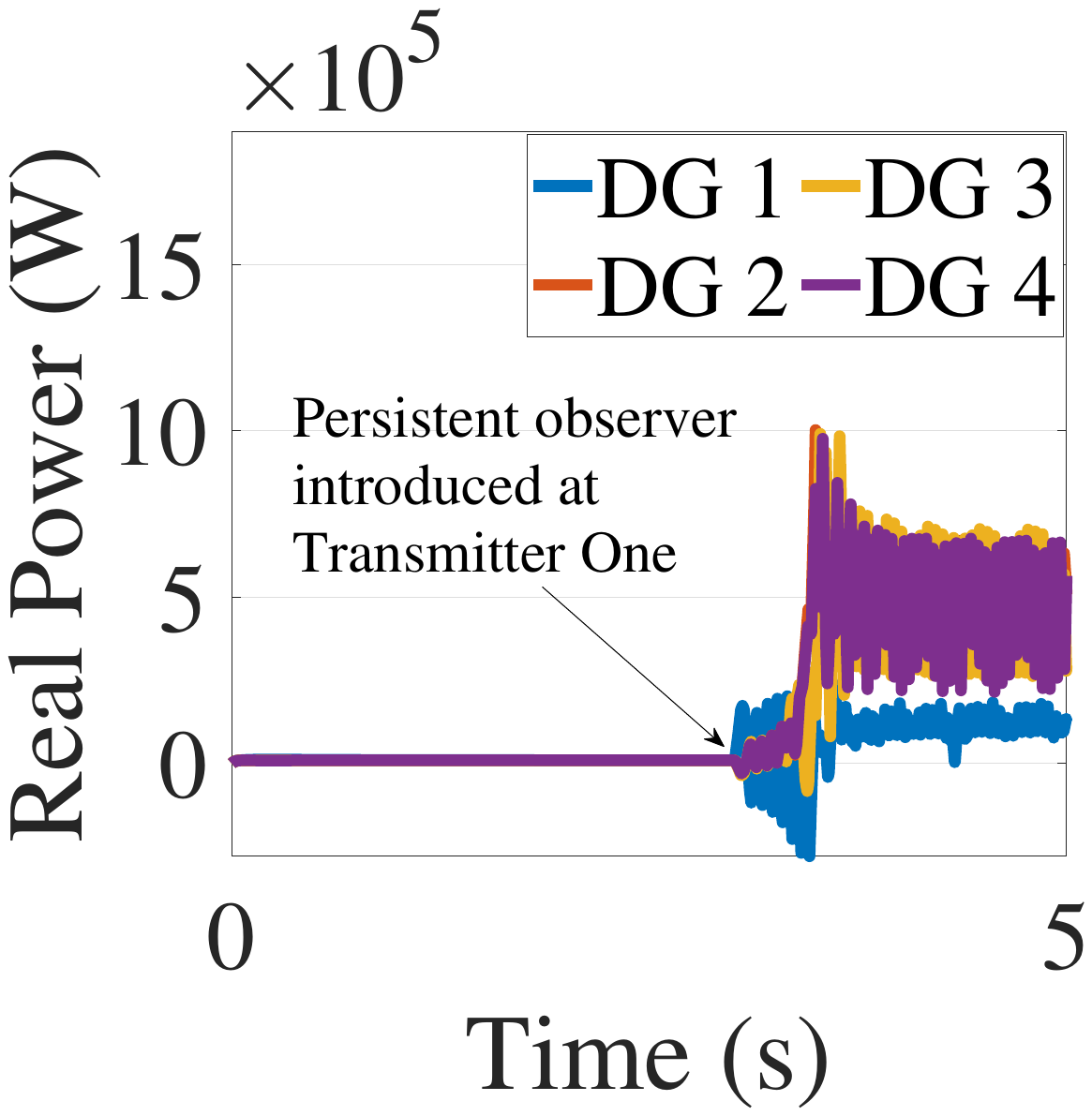}\\[-4ex]
    \includegraphics[width=0.45\linewidth,clip,trim={6 6 6 121}]{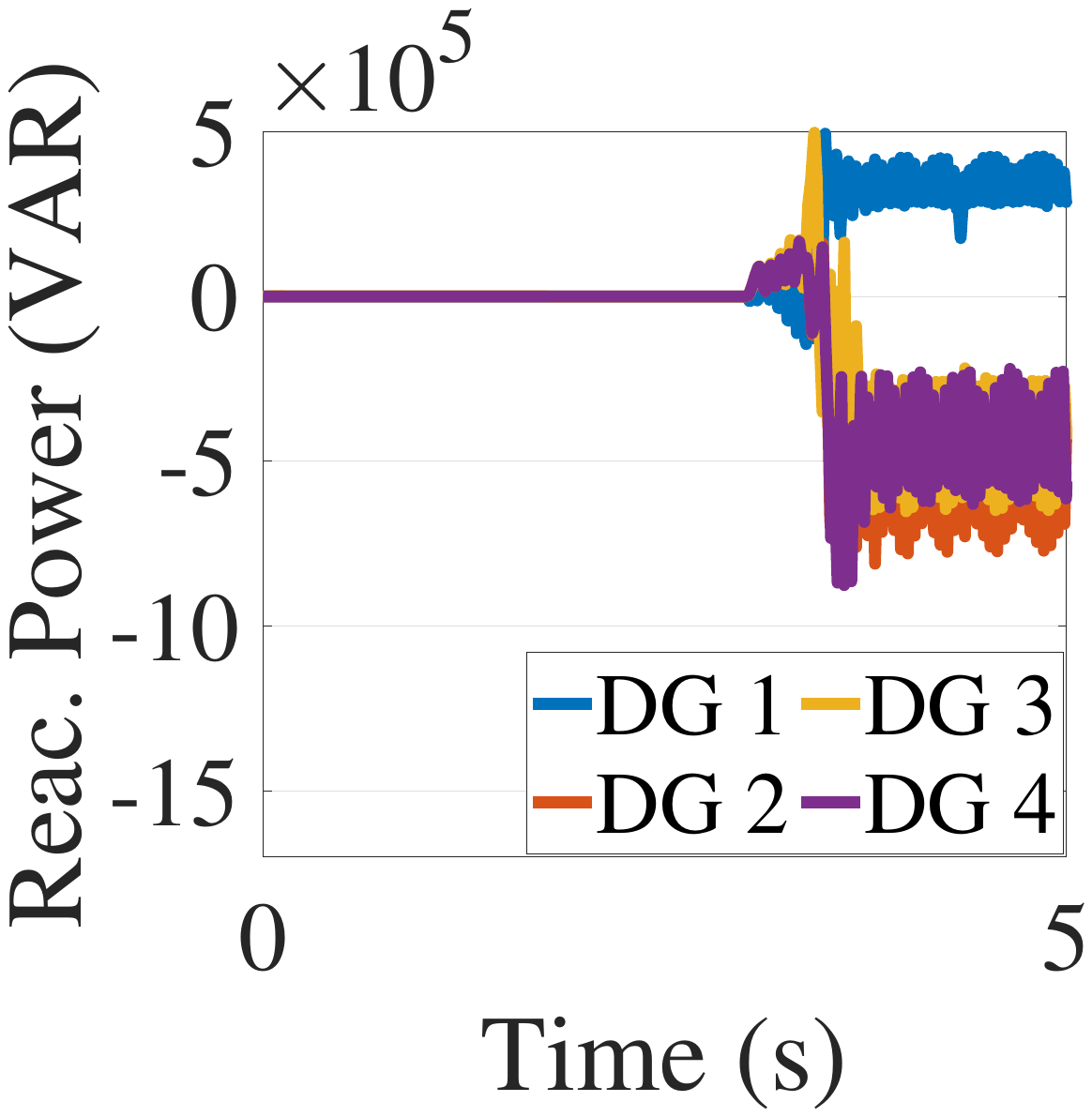}
    \includegraphics[width=0.45\linewidth,clip,trim={6 6 6 137}]{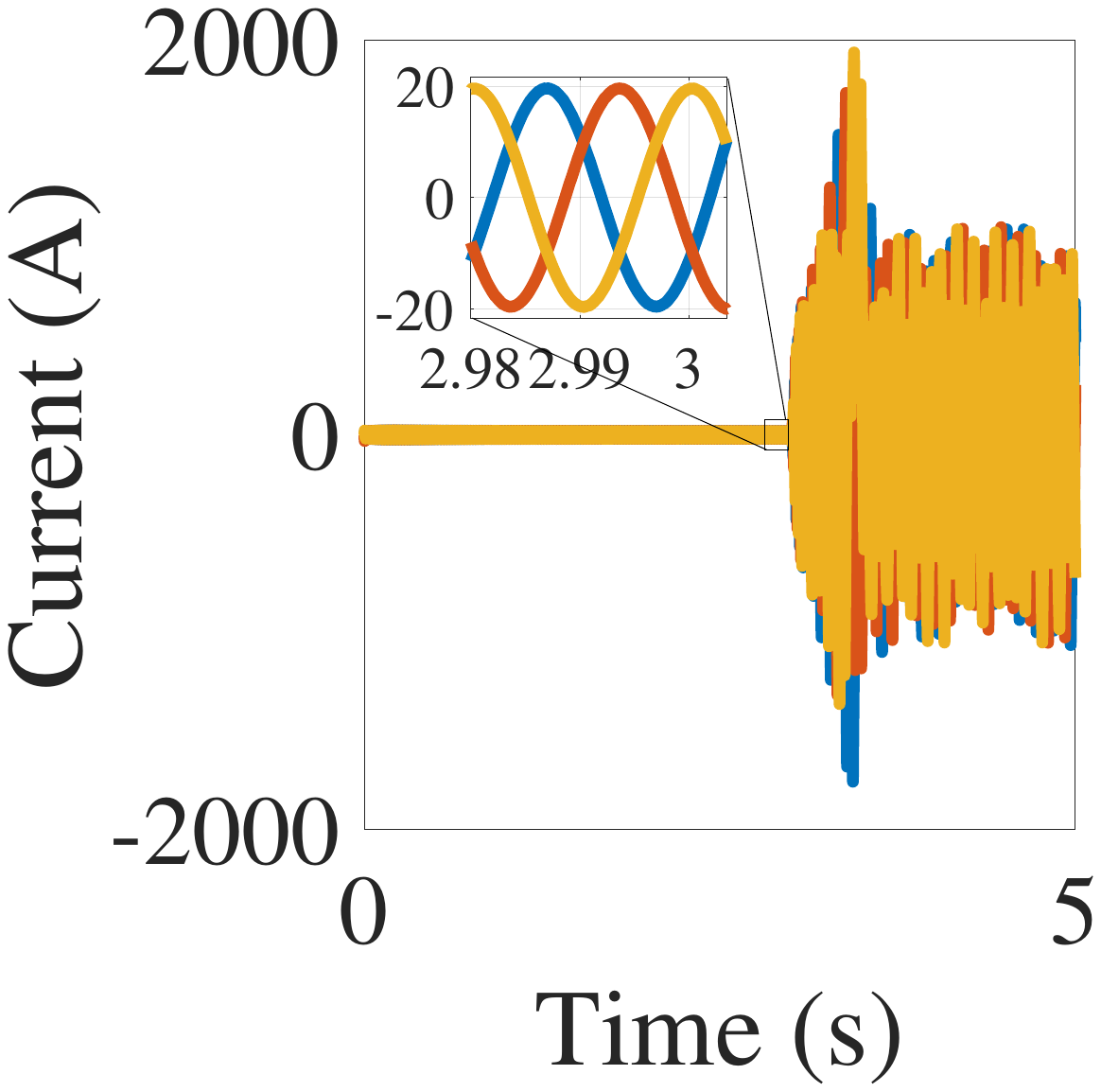}\\[-4ex]
    \includegraphics[width=0.45\linewidth,clip,trim={6 6 6 137}]{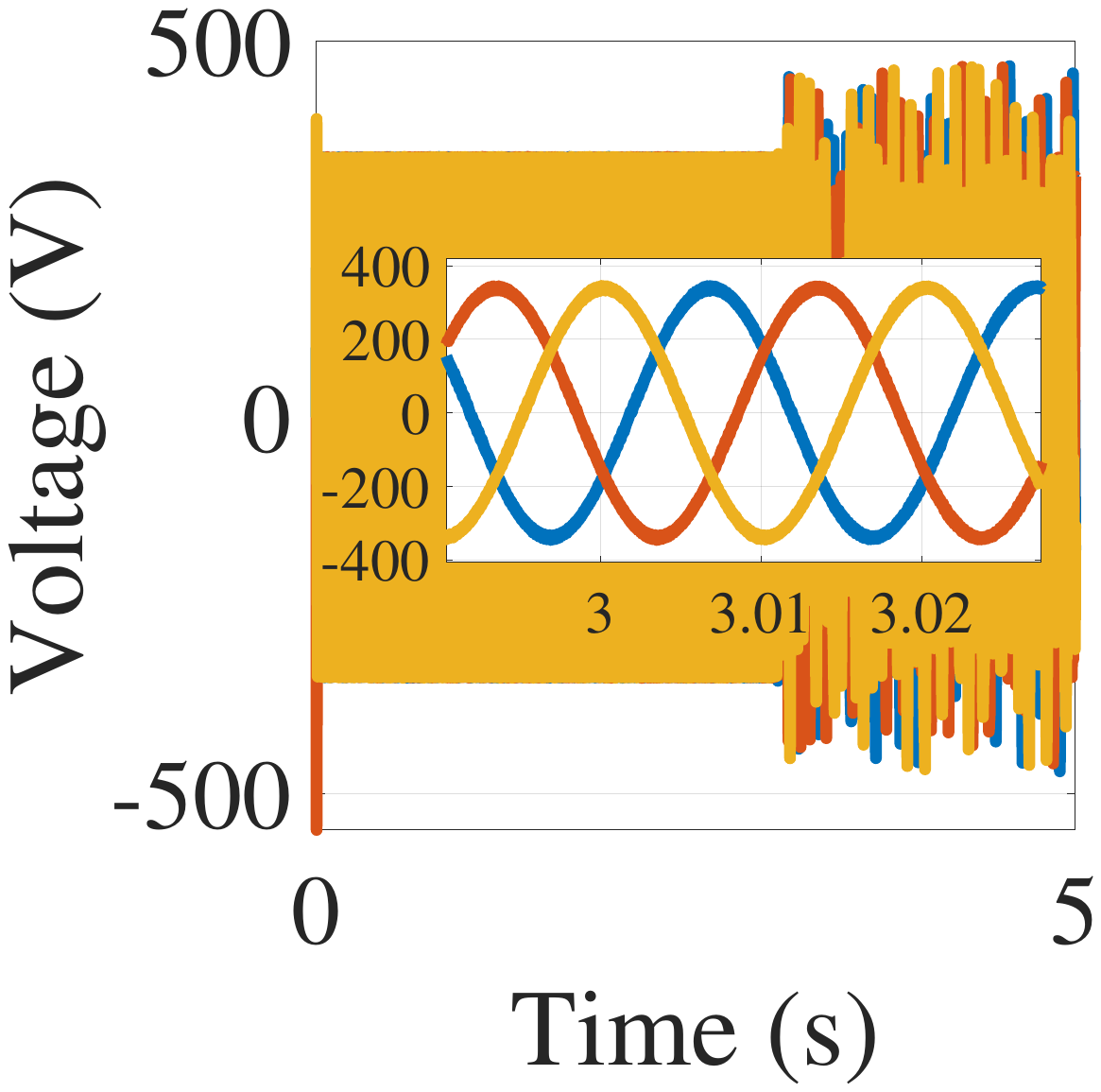}
    \includegraphics[width=0.45\linewidth,clip,trim={6 6 6 137}]{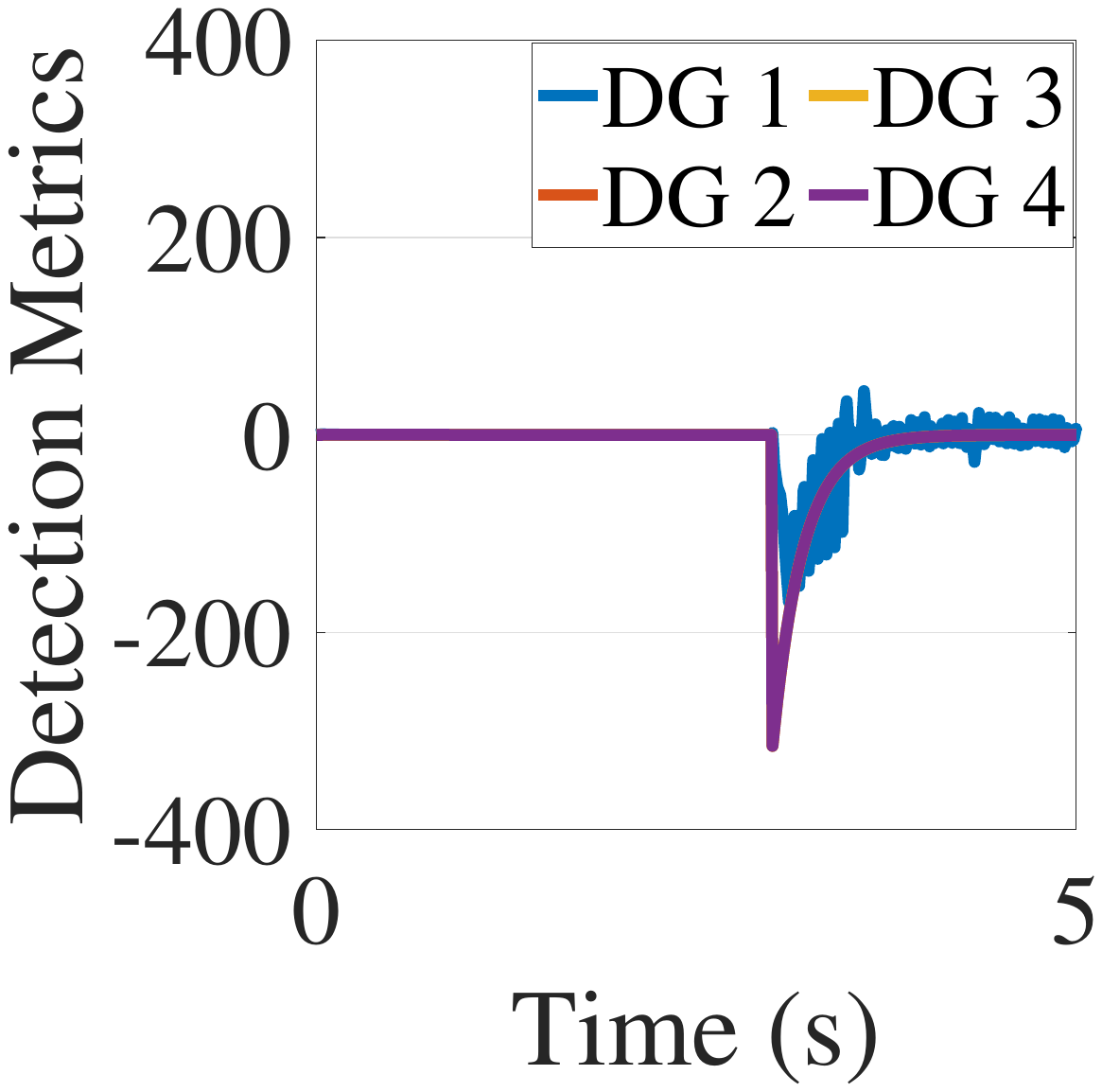}\\[-4ex]
    \caption{{Impact of the presence of a persistent observer (at $t = 3$ s) on a generic QKD-based microgrid.}}
    \label{fig:study2}
\end{figure}


\subsubsection{Short-term Observation} 
In this case study, an attacker attempts to eavesdrop on the quantum channels carrying information from the DG 1 to 2 and 1 to 4 to observe the cryptographic keys as shown in Fig. \ref{fig:QKD} starting from an arbitrarily chosen time instant ($t = 2.32$ s). The attacker fails to gauge the quantum states corresponding to the key due to the Observer effect as explained in \cite{sassoli2013observer}. However, the actions of the attacker create a disturbance of quantum states which is detected by the receiver module using equation \ref{QBER}. This leads to the recipient DG discarding the bad key to prevent incorrect signal decryption. This in turn forces the receiver to discard all keys and corresponding signal values during the period of eavesdropping (observation) effectively resulting in a denial-of-service (DoS) attack. After a few seconds of failed eavesdropping, the attacker stops its attempts at $t = 2.5$ s. This makes the recipient DG able to start receiving accurate measurements from its neighboring DGs again. However, as shown in Fig. \ref{fig:study1} never recuperates from the impact of the short-term observation lasting just $0.18$ s. System parameters show an abnormal trajectory during and after the period of observation. A dip is seen in the frequency trajectory (as measured at DG 1) with sustained disturbances in real and reactive power sharing among DGs. It is seen that system voltage magnitude during the observation (and afterward) continues to increase in an unbounded manner. However, the current magnitude continually decreases as a consequence of the attempted eavesdropping.


\subsubsection{Persistent Observation}
In this setup, the attacker as described in the previous case study chooses to continue its eavesdropping attempts irrespective of whether it can read the quantum keys or not. This leads to a continuous disturbance of the quantum states corresponding to the key and leads to the receiver being unable to decrypt any measurement signals received from the classical communication network. This scenario emulates a continued DoS attack on the intended recipient. Fig. \ref{fig:study2} shows the performance of the quantum microgrid in this scenario.
As a consequence of the persistent observation, system frequency as measured at DG 1 drops by a significant margin. Sustained oscillations are observed in voltage and current magnitudes. Additionally, real and reactive power sharing among DGs is also disturbed.

\begin{figure}
    \centering
    \includegraphics[width=\linewidth]{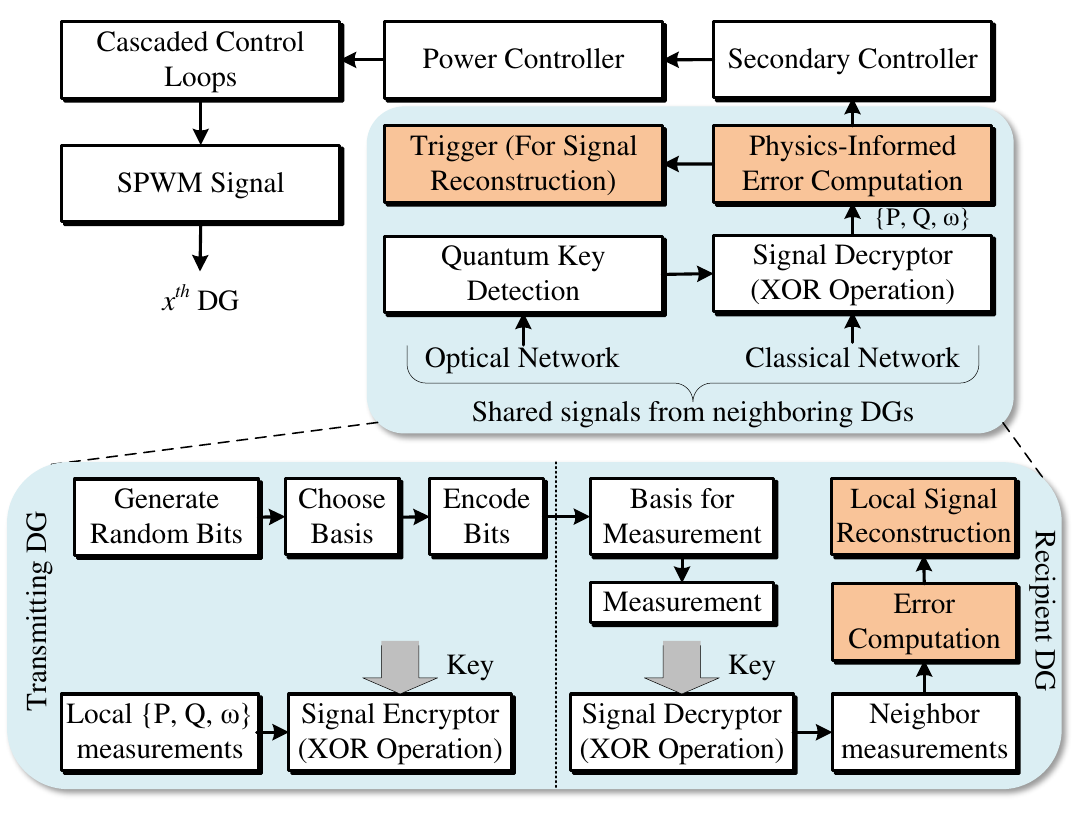}
    \caption{{DG-level error mitigation and control schematics in a microgrid fortified with the proposed QKD framework.}}
    \label{fig:block2}
\end{figure}

\subsection{Fortified Quantum Key Distribution}
For the case studies in this subsection, the proposed fortification framework is integrated with the generic QKD-enabled AC microgrid. The Simulink-based implementation of DG-level error mitigation and control schematics in this framework is depicted in Fig. \ref{fig:block2}. This fortified system is subjected to the same scenarios as depicted in the preceding subsection. This is done to establish a comparative evaluation showcasing the superiority of the proposed detection and mitigation framework over the generic QKD QBER detection and mitigation framework.

\begin{figure}
    \centering
    \includegraphics[width=0.45\linewidth,clip,trim={6 6 6 137}]{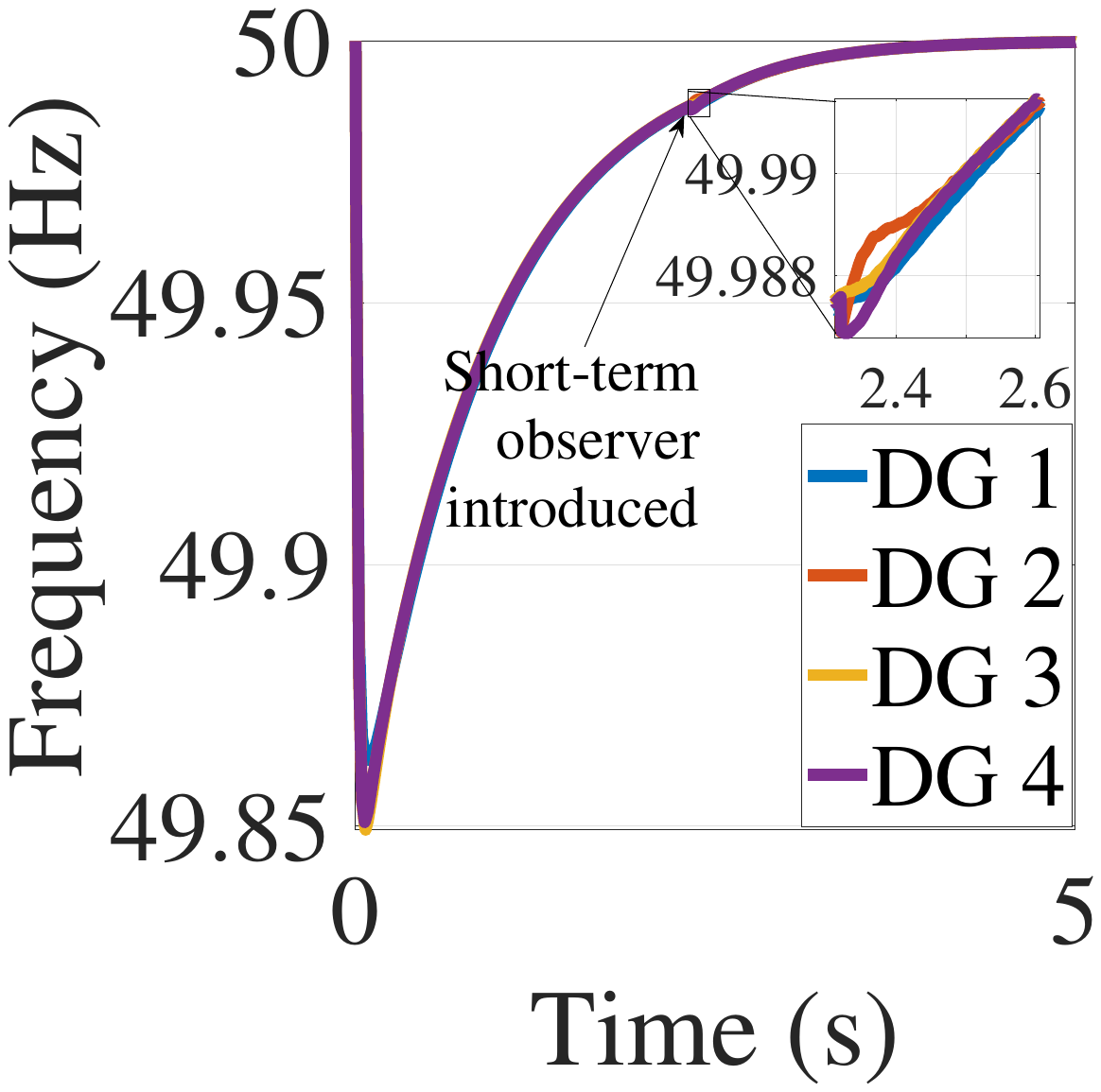}
    \includegraphics[width=0.45\linewidth,clip,trim={6 6 6 121}]{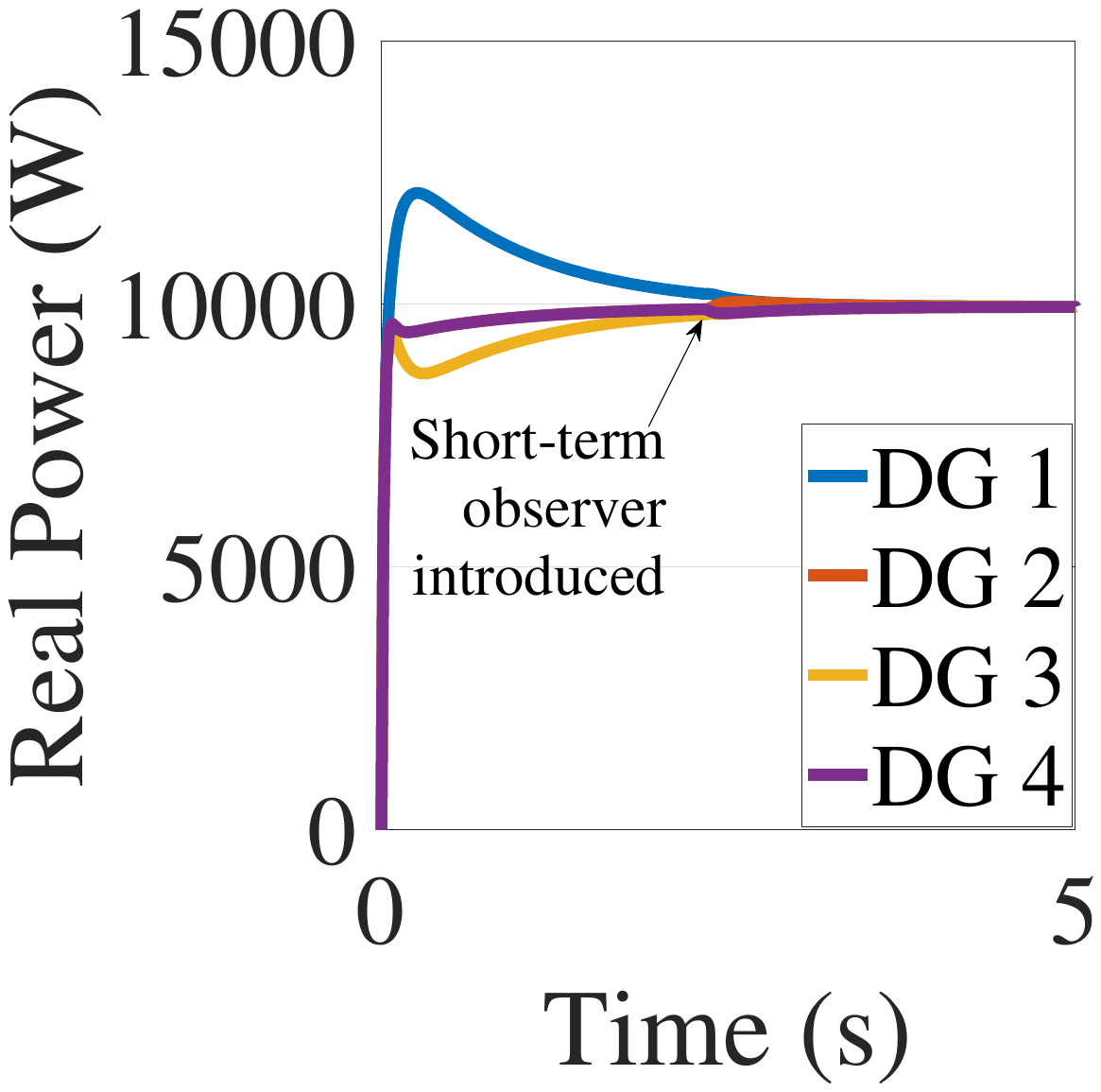}\\[-4ex]
    \includegraphics[width=0.45\linewidth,clip,trim={6 6 6 121}]{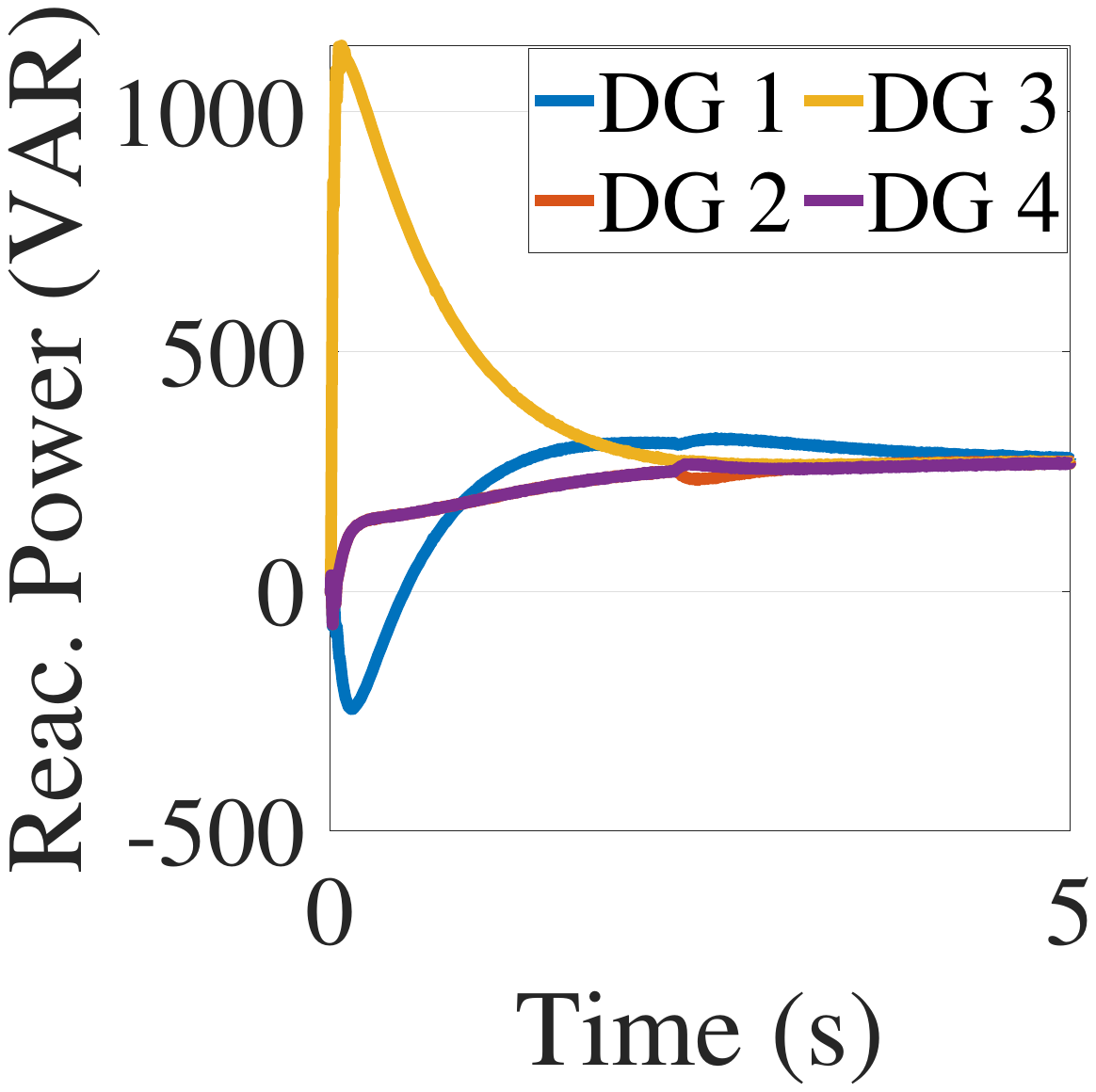}
    \includegraphics[width=0.45\linewidth,clip,trim={6 6 6 137}]{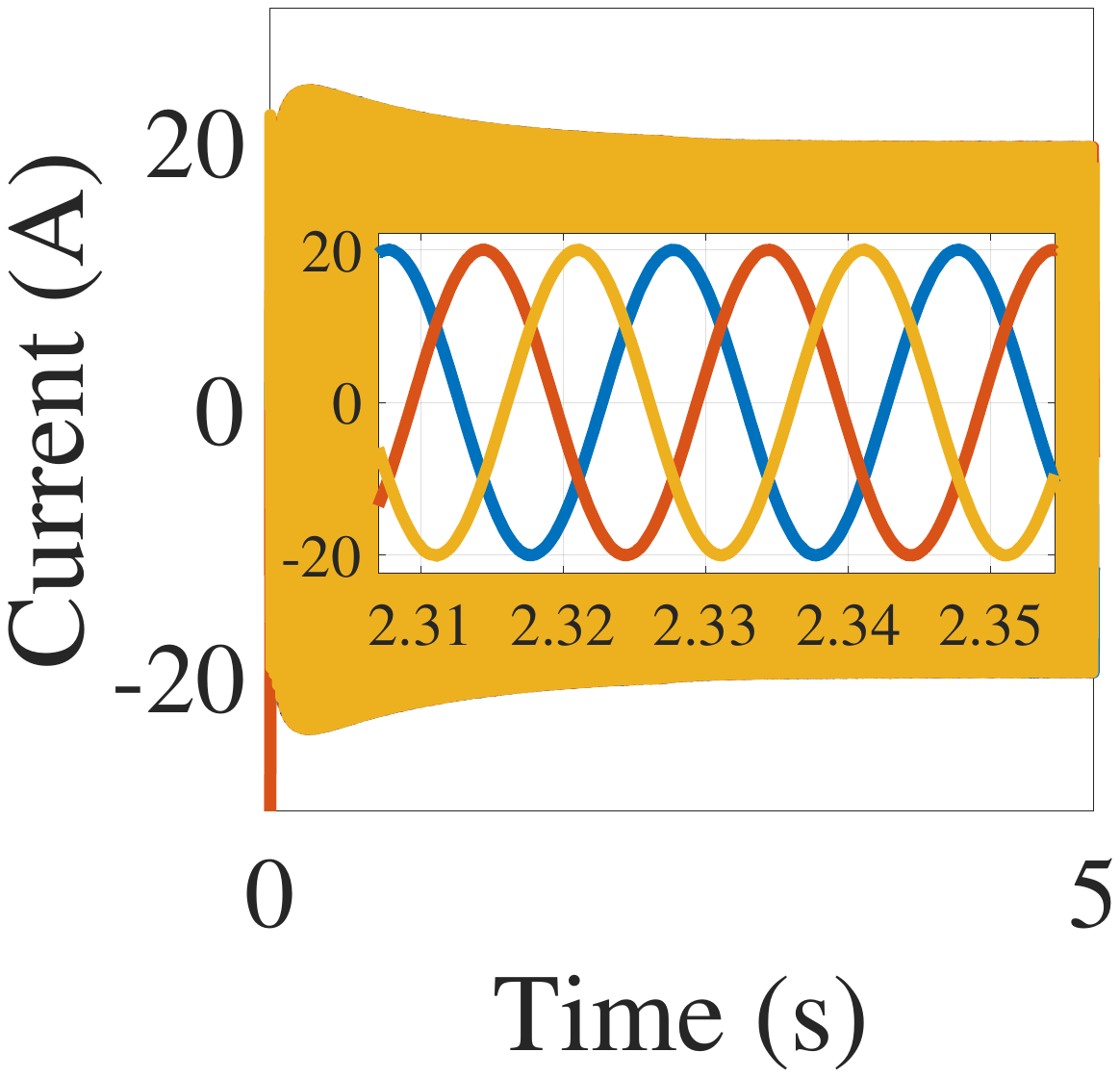}\\[-4ex]
    \includegraphics[width=0.45\linewidth,clip,trim={6 6 6 137}]{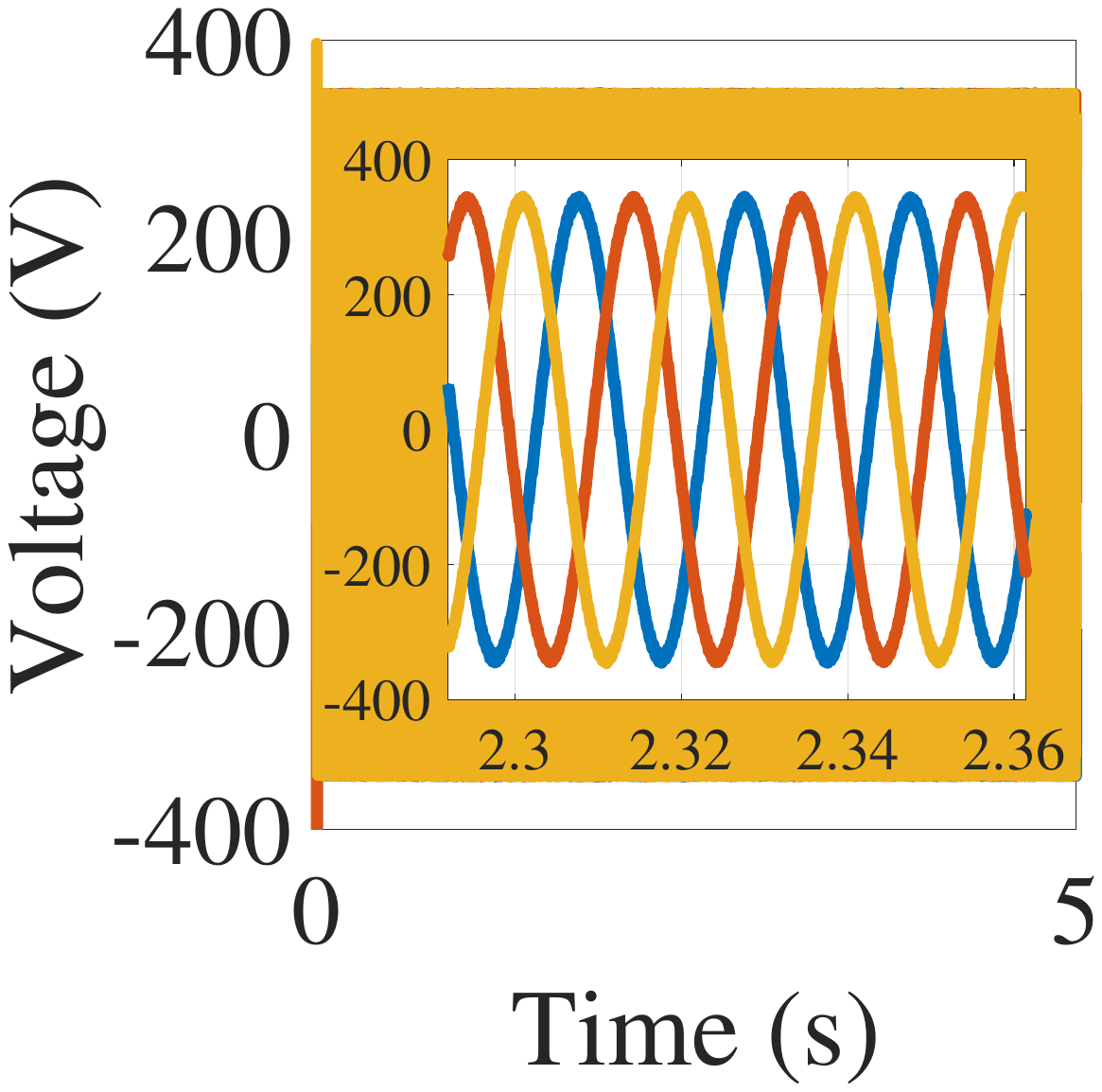}
    \includegraphics[width=0.45\linewidth,clip,trim={6 6 6 137}]{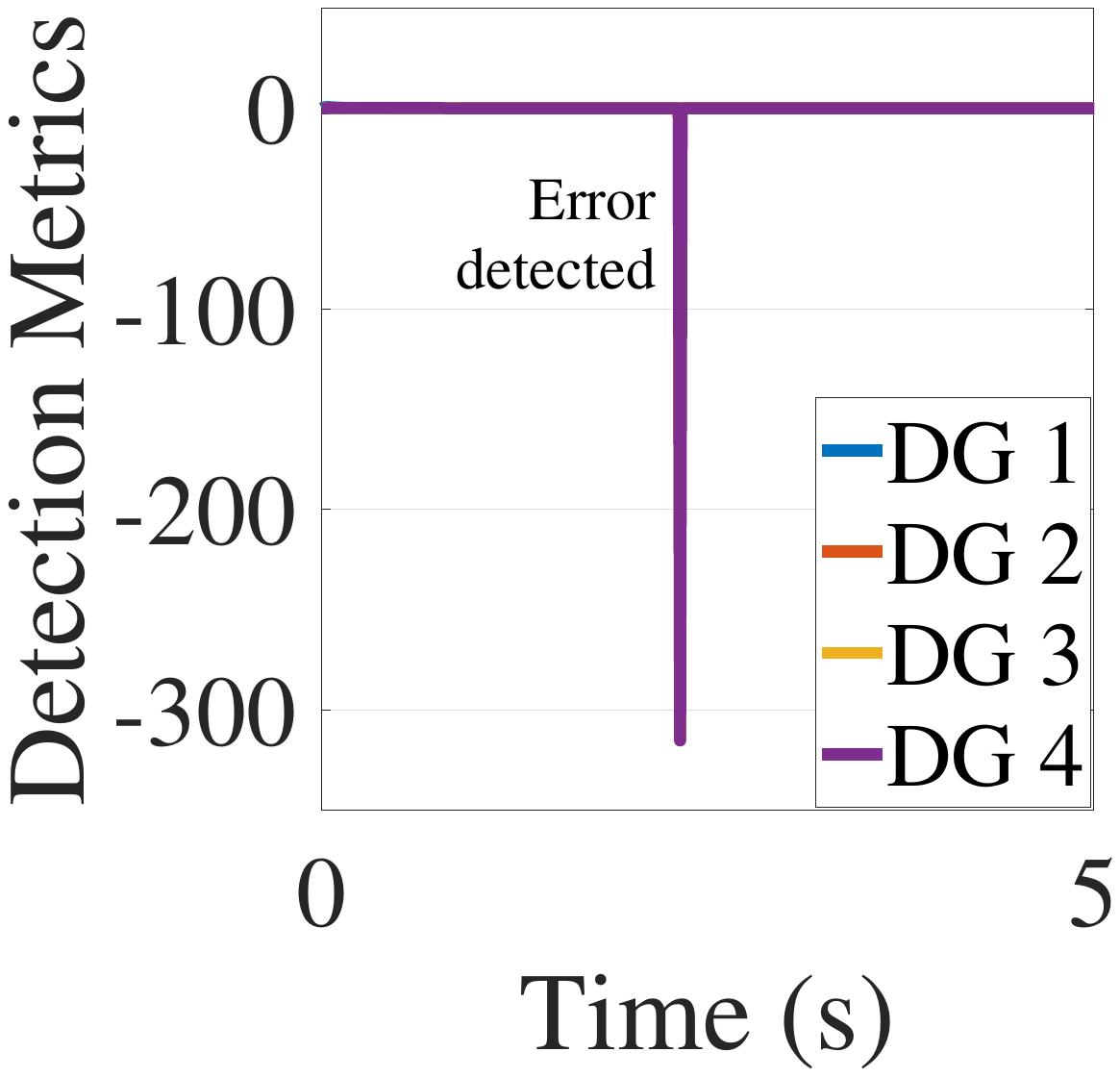}\\[-4ex]
    \caption{{Performance of the proposed QKD-based, fortified microgrid model in the presence of a short-term observer.}}
    \label{fig:study4}
    \vspace{-10pt}
\end{figure}

\subsubsection{Resilience against Short-term Observation}
In this case study, the attacker tries to eavesdrop on the quantum channels connecting DG 1 to 2 and 1 to 4. The period of eavesdropping (observation) is kept identical to section V-A-1. In this setting, QBER detection is performed via the detection metrics in Table \ref{tab:2}. Fig. \ref{fig:study4} shows the performance of the fortified QKD-enabled microgrid in this scenario. The spike observed in the trajectory of the detection metrics ($DM^k_1+DM^k_2$) associated with the recipient DG 
denotes a high QBER triggering the reconfiguration of the active adjacency matrix as per the steps in section IV-B. 
The microgrid state trajectory showcases that the proposed fortification mechanism protects the system against quantum state disturbances caused by the attacker's (eavesdropper's) actions.


\begin{figure}
    \centering
    \includegraphics[width=0.45\linewidth,clip,trim={6 6 6 137}]{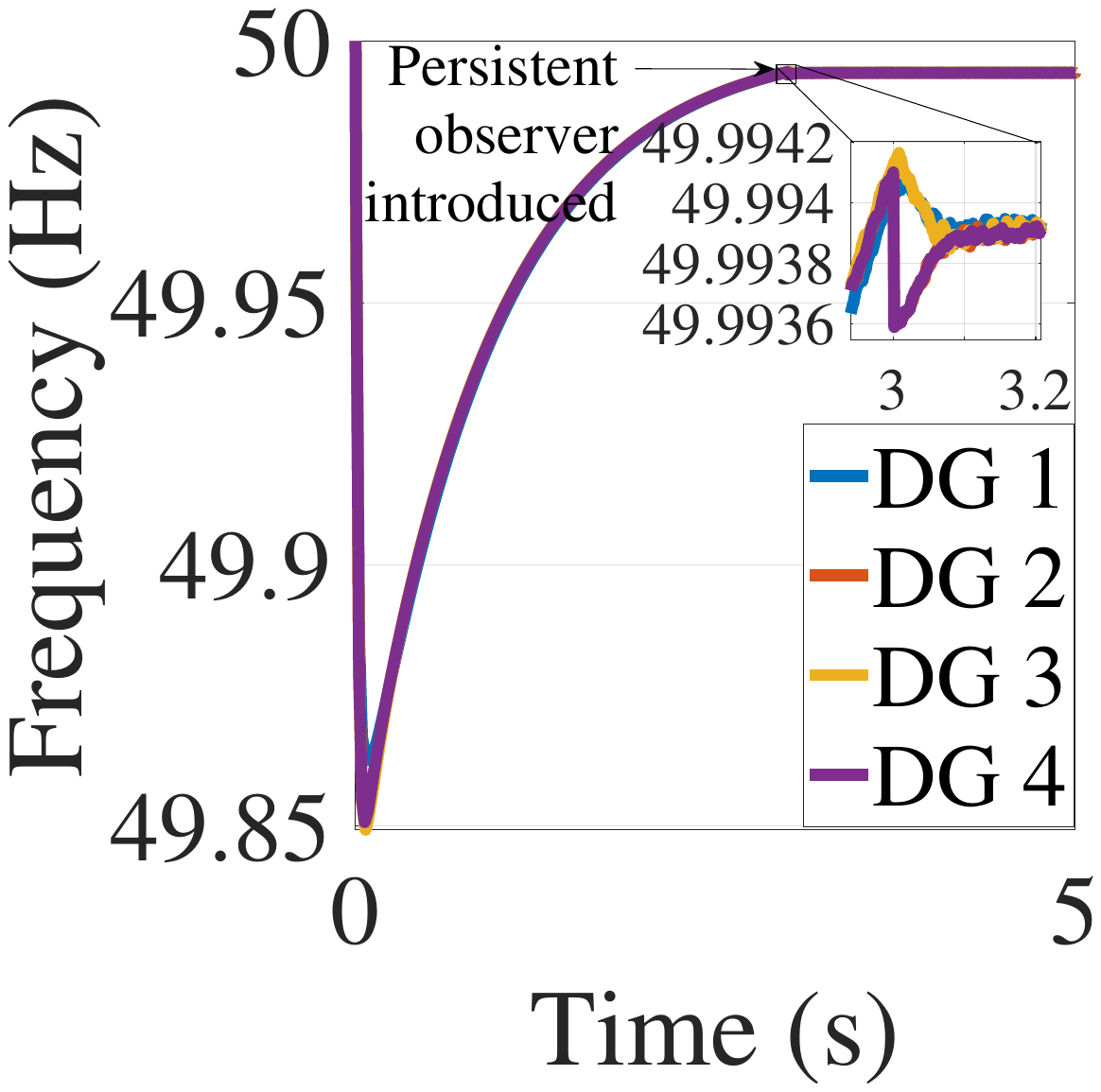}
    \includegraphics[width=0.45\linewidth,clip,trim={6 6 6 121}]{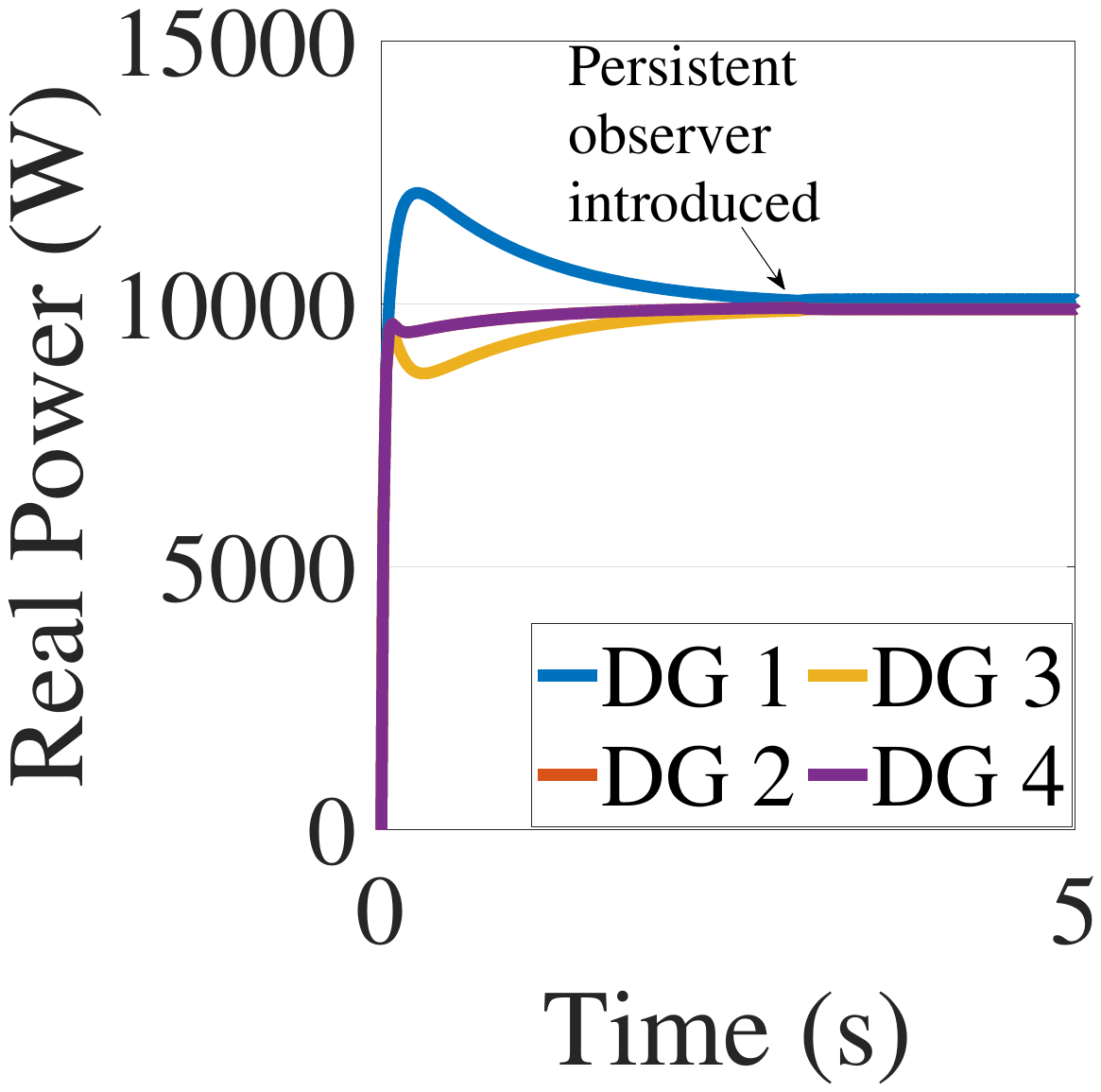}\\[-4ex]
    \includegraphics[width=0.45\linewidth,clip,trim={6 6 6 121}]{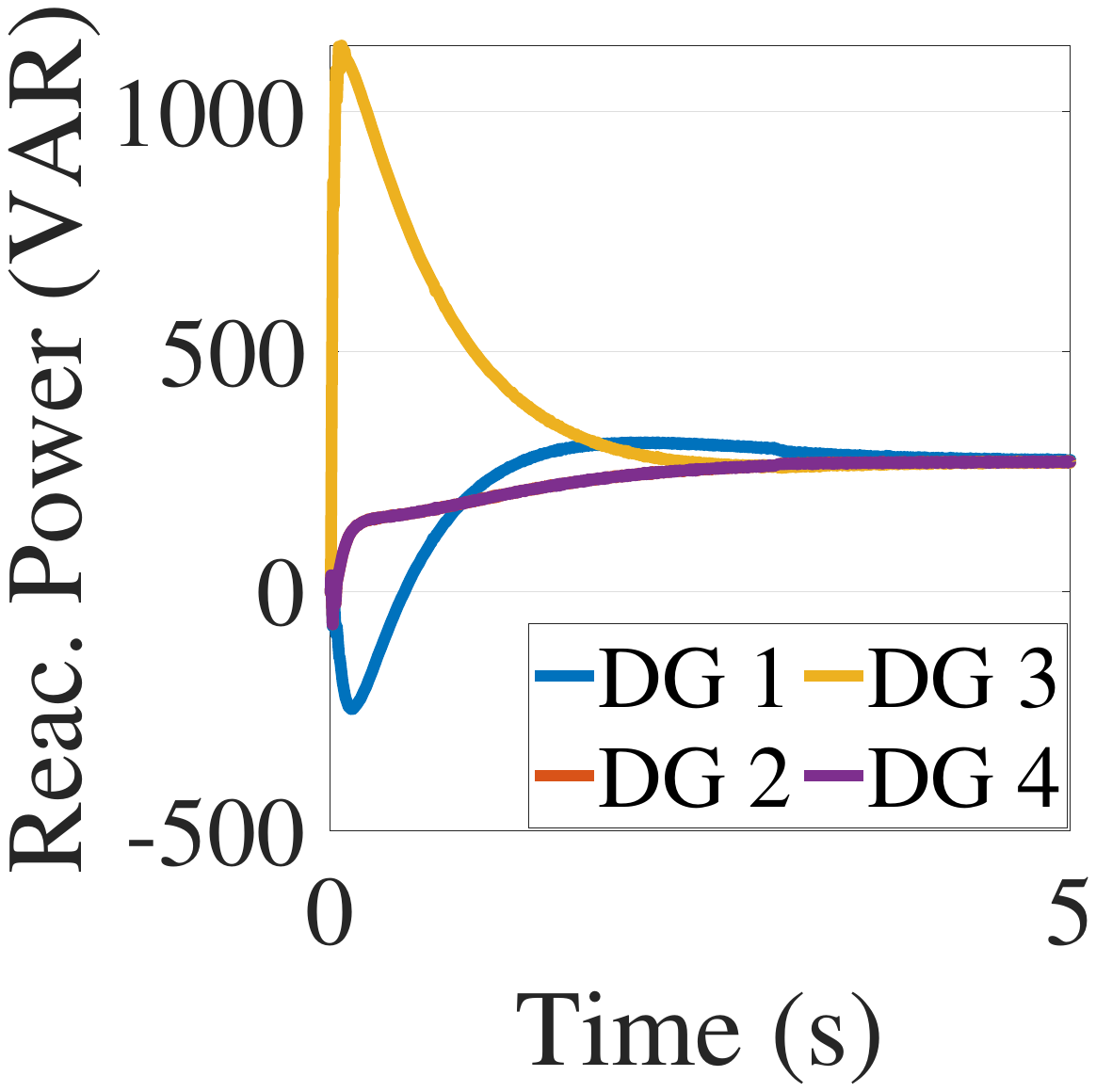}
    \includegraphics[width=0.45\linewidth,clip,trim={6 6 6 137}]{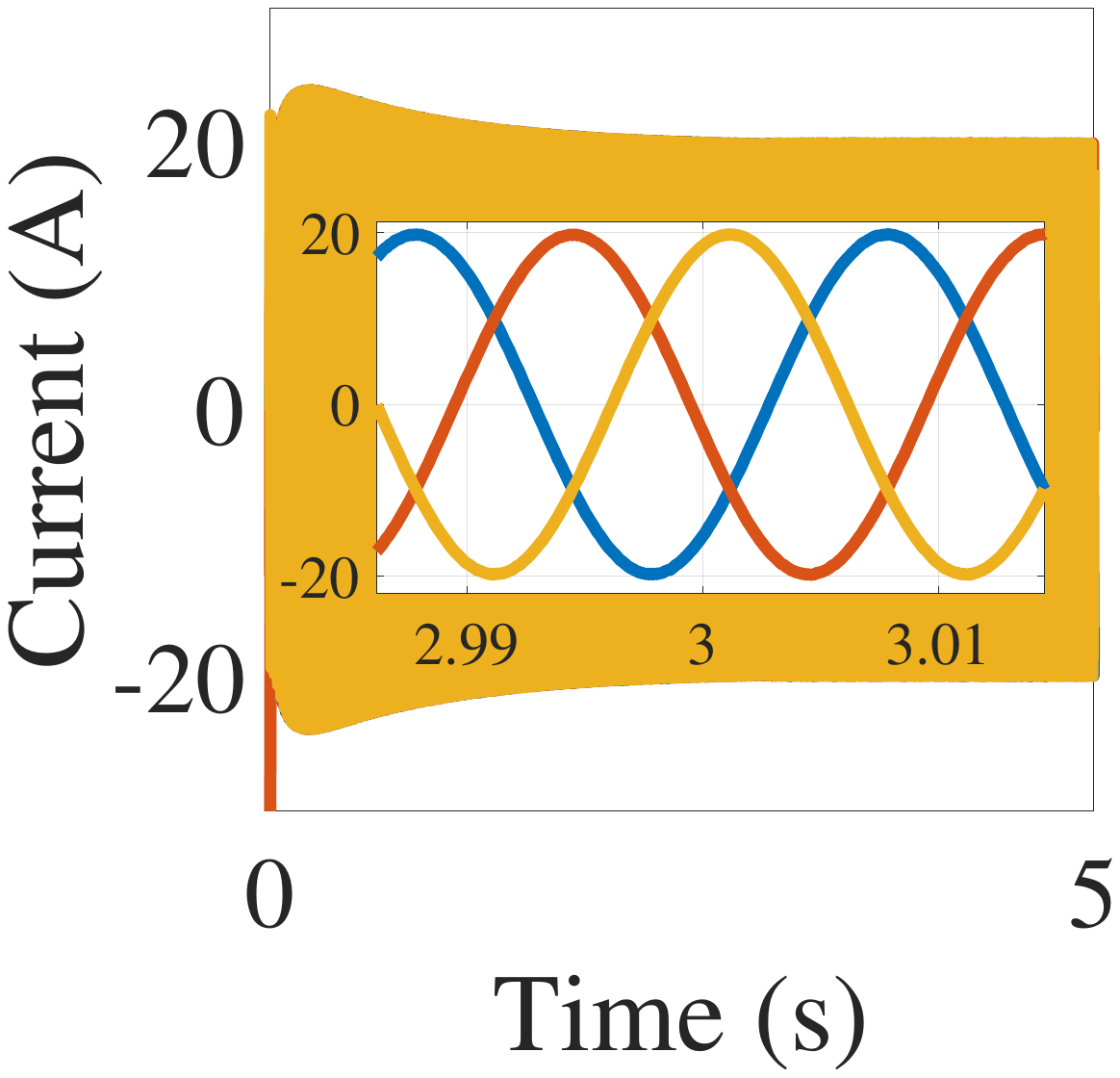}\\[-4ex]
    \includegraphics[width=0.45\linewidth,clip,trim={6 6 6 137}]{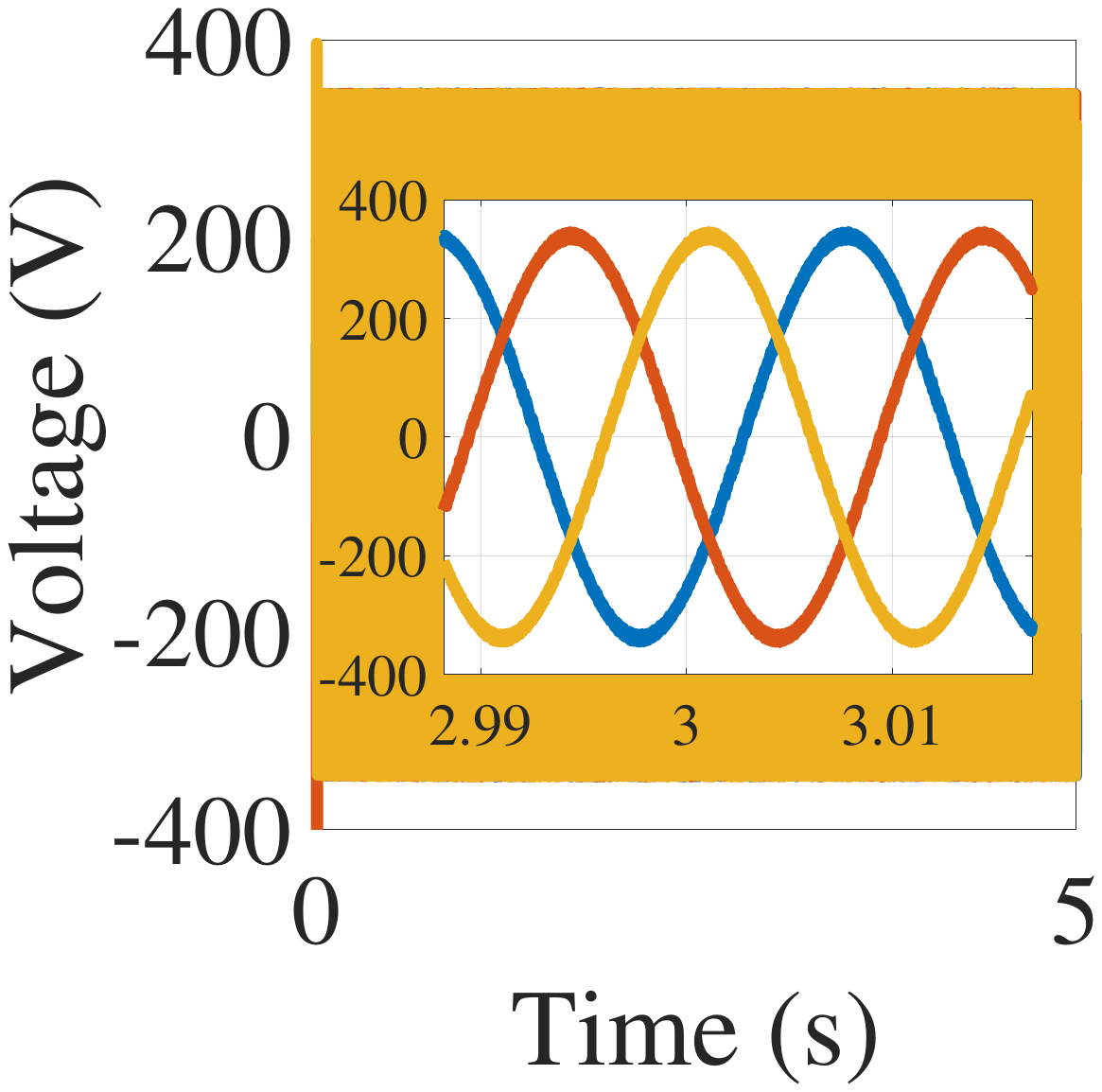}
    \includegraphics[width=0.45\linewidth,clip,trim={6 6 6 137}]{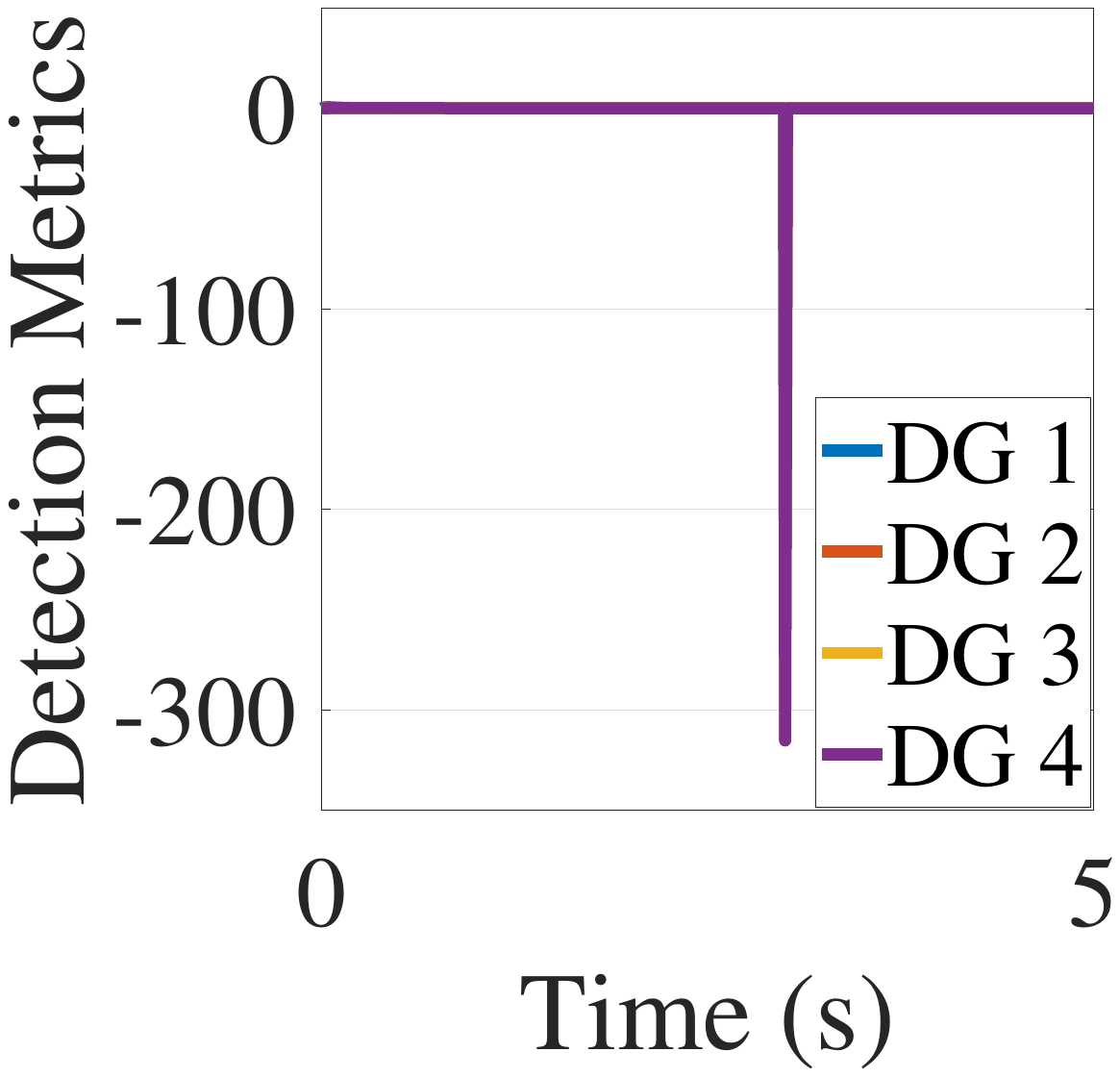}\\[-4ex]
    \caption{{Impact of a persistent observer on a QKD-based microgrid fortified with the proposed signal reconstruction framework.}}\vspace{-1ex}
    \label{fig:study5}
\end{figure}



\subsubsection{Resilience against Persistent Observation}
In this case, the quantum channels carrying keys from DGs 1 to 2 and 1 to 4 in the fortified quantum microgrid are subjected to an observation period that begins at $t = 3$ s and continues till the end of the simulation period. As shown in  Fig. \ref{fig:study5}, the presence of the persistent eavesdropper does not affect system dynamics in an observable manner as the system reconfigures its active adjacency matrix, isolating the communication links between DGs 1 and 2.

\subsubsection{Resilience against ($N-1$) Manipulation}
For this case study, an attacker with access to DGs 2, 3, and 4 is introduced into the fortified QKD-enabled microgrid environment at $t = 2$ s. The attacker in this scenario, performs simultaneous manipulation of all measurements emanating from the infected DGs.
As shown in Fig. \ref{steps}, elements of the error propagation matrix are identified via the spikes in detection metric magnitudes ($DM^k_1+DM^k_2$). To counter this attack strategy, the proposed fortification strategy formulates an intermediate matrix $S_3$ that adheres to equation (\ref{mitigation}).
Matrix $S_3$ ensures that the manipulated measurements in the QKD-enabled microgrid environment are replaced with {reconstructed} measurements. Once the \texttt{False} measurements are eliminated from the system, the framework switches to the default topology $S_1$. Fig. \ref{steps} shows that the proposed fortification framework effectively protects the QKD-enabled microgrid against adversarial node-level manipulations. As depicted in the figure all system parameters continue to follow a stable nominal trajectory even during the attack period.

\section{Conclusion}
% ...
State dynamics within the QKD-enabled microgrid environment can be destabilized due to the presence of eavesdroppers. This is caused due to the observer effect where any attempt to observe the state of the key in a quantum channel disrupts it and incorporates errors. Errors in quantum keys can lead to incorrect message decryptions resulting in flawed control signals. Another major vulnerability in QKD-enabled microgrids arises when attackers with access to DG-level devices manipulate measurement signals before encryption. This leads to the attack propagating to the control layer (without being detected at the quantum layer) ultimately incorporating erroneous values in the microgrid state trajectory.

To address these problems, this paper combines the conventional security features of QKD with localized, DG-level detection metrics that identify and flag the disturbance in quantum keys generated due to observation attempts by external entities. The proposed strategy detects and isolates erroneous inputs at the receiver level by modifying the active adjacency matrix as per which information flow is realized in the microgrid network. This strategy serves a dual purpose by excluding communication links where eavesdroppers may have access to network devices along with incoming signals from infected and flagged DGs. It is observed that the proposed framework can protect the system even when $(N-1)$ DGs are manipulated simultaneously. Future extension of this project will seek to enhance the presented strategy by achieving full resiliency.

\ifCLASSOPTIONcaptionsoff
  \newpage
\fi
\bibliographystyle{IEEEtran}
\bibliography{ref}

\end{document}